# *Spontaneous Rotational Symmetry Breaking in a Kramers Two-Level System*

*Mário G. Silveirinha*[*]

[1] *University of Lisbon–Instituto Superior Técnico and Instituto de Telecomunicações, Avenida Rovisco Pais, 1, 1049-001 Lisboa, Portugal, mario.silveirinha@co.it.pt*

**Abstract**

Here, I develop a model for a two-level system that respects the time-reversal symmetry of the atom Hamiltonian and the Kramers theorem. The two-level system is formed by two Kramers pairs of excited and ground states. It is shown that due to the spin-orbit interaction it is in general impossible to find a basis of atomic states for which the crossed transition dipole moment vanishes. The parametric electric polarizability of the Kramers two-level system for a definite ground-state is generically nonreciprocal. I apply the developed formalism to study Casimir-Polder forces and torques when the two-level system is placed nearby either a reciprocal or a nonreciprocal substrate. In particular, I investigate the stable equilibrium orientation of the two-level system when both the atom and the reciprocal substrate have symmetry of revolution about some axis. Surprisingly, it is found that when chiral-type dipole transitions are dominant the stable ground state is not the one in which the symmetry axes of the atom and substrate are aligned. The reason is that the rotational symmetry may be spontaneously broken by the quantum vacuum fluctuations, so that the ground state has less symmetry than the system itself.

---

[*] To whom correspondence should be addressed: E-mail: *mario.silveirinha@co.it.pt*

# I. Introduction

At the microscopic level, physical systems are generically ruled by time-reversal invariant Hamiltonians [1]. The time reversal operator $\mathcal{T}$ acts on the system state in such a way that its dynamics is effectively reversed in time, analogous to a movie played backwards [1, 2]. Evidently, a double time-reversal should essentially undo the action of $\mathcal{T}$, i.e., it must bring the system back to its original state. This could naively suggest that $\mathcal{T}^2$ should be the identity operator ($\mathcal{T}^2 = \mathbf{1}$). Notably, this is not always the case. For example, for spin ½ particles, e.g., an electron described by the Schrödinger equation, the time reversal operator satisfies $\mathcal{T}^2 = -\mathbf{1}$ [3]. The extra minus sign implies a change in the phase of the wave function but does not alter the expectation of any physical operator, consistent with the idea that the double time reversal should leave the system state unchanged.

As is well known, the extra minus sign has several interesting physical consequences. For example, the wave scattering in time-reversal invariant platforms with $\mathcal{T}^2 = -\mathbf{1}$ is characterized by an anti-symmetric scattering matrix [4, 5]. This property may enable propagation immune to the back-scattering due to impurities or deformations of the propagation path, a phenomenon known as the spin-Hall effect [4]. Furthermore, another nontrival implication of $\mathcal{T}^2 = -\mathbf{1}$ is that the stationary states of an electronic (spin ½) system must be doubly degenerate. This property is known as the Kramers theorem [3].

The archetypal system in quantum optics is the two-level atom. In the usual description, the atom has two *non-degenerate* energy states: the excited state and the ground state. Evidently, such an idealized model is at odds with the Kramers theorem, as the time reversal invariance of the atom Hamiltonian requires both the excited state and the ground state to be doubly degenerate. In fact, a time-reversal invariant spin ½ system with two different energy levels has necessarily *four* eigenstates.

The main objective of this article is to understand how the degeneracy of the ground state and the additional complexity arising from the dipolar-type interactions between the two ground states and the two excited states of the Kramers two-level system affects Casimir-Polder interactions and the stable ground state configuration [6-9]. To this end, I calculate the Casimir interaction energy (van der Waals potential) of the Kramers two-level atom both in reciprocal and in nonreciprocal environments. I find that for nonreciprocal environments the Casimir energy is ground state dependent and that the interactions with the quantum vacuum may result in an energy splitting of the free-ground eigenstates, somewhat alike to the Zeeman effect. It is important to mention that several recent works highlighted that the nonreciprocity of the environment can tailor in unique ways the Casimir-Polder dispersion forces and torques in atomic and nano-scale systems [10-16], but neglecting in most cases the degeneracy of the atom ground state.

In addition, it is proven that for reciprocal environments the Casimir energy is independent of the ground state. I use the developed formalism to study the equilibrium configurations of a Kramers two-level system positioned near a metallic surface. The atom is invariant under arbitrary rotations about some symmetry axis. Intuitively, one might expect that the stable ground state configuration should have rotational symmetry, such that the symmetry axes of the atom and of the substrate should be aligned. Interestingly, it is found that this intuitive understanding is wrong and that the equilibrium configuration depends on the relative strength of the dipolar transitions with circular and linear polarization. When the chiral-type (with circular polarization) dipolar transitions dominate, the stable ground state has a broken rotational symmetry and the atom symmetry axis is parallel to the metal substrate. Thus, the ground state has less symmetry than the system itself: the rotational symmetry is spontaneously broken by the quantum vacuum fluctuations.

In general, the mechanism of spontaneous symmetry breaking plays a fundamental role in many physical phenomena, e.g., in ferroelectricity, ferromagnetism or in superconductivity [17, 18, 19]. The Higgs mechanism is another example of spontaneous symmetry breaking and explains the origin of the mass of bosonic particles in the standard model of elementary particle physics [19]. The spontaneous symmetry breaking mechanism is also relevant in the context of the electrodynamics of moving media and quantum friction [20, 21]. Here, I show for the first time that the interactions of a Kramers two-level system with the quantum vacuum may result in a spontaneously broken rotational-symmetry.

## II. Two-level system formed by Kramers pairs

According to the Kramers theorem, the stationary states of a spin ½ electronic system described by a time-reversal invariant Hamiltonian must be degenerate [3]. Indeed, given a generic state $\left|n_1\right\rangle$ with energy $E_n$ it is possible to construct another state, $\left|n_2\right\rangle = \mathcal{T}\left|n_1\right\rangle$, with the same energy $E_n$. The two states are necessarily linearly independent because for spin ½ systems the time reversal operator satisfies $\mathcal{T}^2 = -\mathbf{1}$. In particular, it follows that the ground state of any system with a single electron is necessarily degenerate. For example, the ground state of the hydrogen atom is determined by the *s*-orbital which consists of two states with opposite spins. In a system with a degenerate ground the time-reversal operator typically links different ground states.

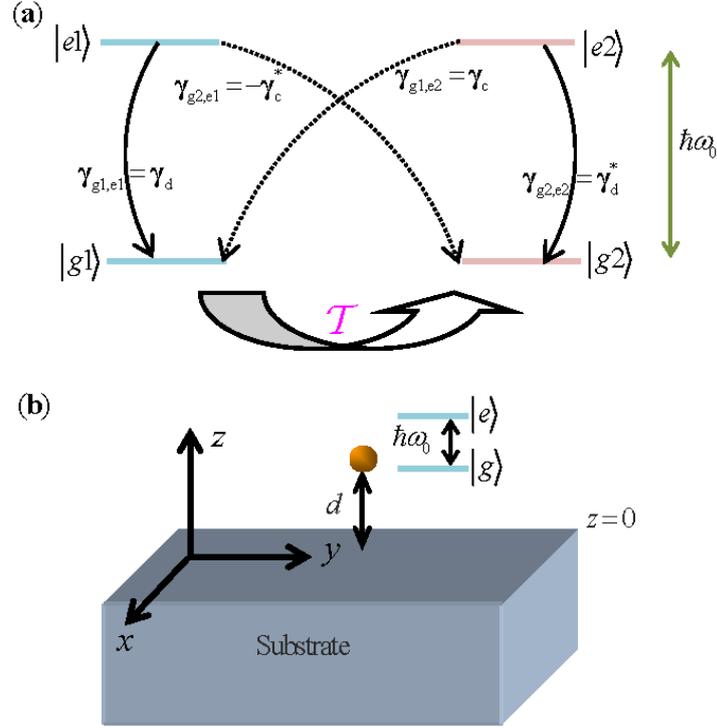

**Fig. 1 a)** A two-level system is formed by a minimum of four eigenstates, linked in pairs by the time reversal operator $\mathcal{T}$ (Kramers pairs). The black arrows indicate the nontrivial dipole moment downward transitions ($\gamma_{g,e}$). The upward transition elements are $\gamma_{e,g} = \gamma_{g,e}^*$ (not shown in the figure). **b)** A two-level system formed by two Kramers pairs is placed at a distance $d$ from a planar material substrate.

From the Kramers theorem, it follows that the minimal basis for a $\mathcal{T}$-invariant two-level system consists of two excited states and two ground states (Fig. 1a), rather than a single excited state and a single ground state as considered for simplicity in most works. Let $|e_1\rangle$, $|e_2\rangle$ be a basis of the excited states and $|g_1\rangle$, $|g_2\rangle$ a basis of the ground states. The two sets are formed by Kramers pairs such that

$$|e_2\rangle = \mathcal{T}|e_1\rangle, \qquad |g_2\rangle = \mathcal{T}|g_1\rangle. \tag{1}$$

From $\mathcal{T}^2 = -\mathbf{1}$ it follows that $|e_1\rangle = -\mathcal{T}|e_2\rangle$ and $|g_1\rangle = -\mathcal{T}|g_2\rangle$.

The atom Hamiltonian is $\hat{H}_{at} = \sum_i E_i |i\rangle\langle i|$ ($i = e_1, e_2, g_1, g_2$). The atomic transition frequency is $\omega_0 = (E_e - E_g)/\hbar$ with $E_e \equiv E_{e1} = E_{e2}$ and $E_g \equiv E_{g1} = E_{g2}$. The light-matter interactions are determined by the direct coupling Hamiltonian (electric dipole approximation) $\hat{H}_{int} = -\hat{\mathbf{p}} \cdot \mathbf{E}$, with $\mathbf{E}$ the electric field and $\hat{\mathbf{p}}$ the atomic dipole moment operator. For a two-level atom with four stationary states the dipole moment operator can be written as $\hat{\mathbf{p}} = \sum_{i,j} \boldsymbol{\gamma}_{ij} \sigma_{ij}$ with $\sigma_{ij} = |i\rangle\langle j|$ and $\boldsymbol{\gamma}_{ij} = \langle i|\hat{\mathbf{p}}|j\rangle = \boldsymbol{\gamma}_{ji}^*$ the dipole moment matrix elements in the considered basis ($i, j = e_1, e_2, g_1, g_2$).

The time-reversal operator for the one-body Schrödinger equation is of the form $\mathcal{T} = \mathcal{U}\mathcal{K}$, with $\mathcal{K}$ the complex conjugation operator and $\mathcal{U} = i\boldsymbol{\sigma}_y$ determined by the Pauli matrix $\boldsymbol{\sigma}_y$ [3]. Taking into account that $\mathcal{T}$ is anti-linear and $\mathcal{T}^{-1} = -\mathcal{T}$, it can be shown that for a generic time-reversal invariant operator $\hat{A}$ ($\hat{A} = \mathcal{T}^{-1}\hat{A}\mathcal{T}$ with $\hat{A}$ Hermitian) one has

$$\langle i|\hat{A}|j\rangle = \langle \mathcal{T}i|\hat{A}|\mathcal{T}j\rangle^*, \qquad (2)$$

for generic states $|i\rangle$ and $|j\rangle$. Replacing $|i\rangle \to |\psi\rangle$ and $|j\rangle \to |\mathcal{T}\psi\rangle$ it follows that $\langle \psi|\hat{A}|\mathcal{T}\psi\rangle = 0$ for an arbitrary $|\psi\rangle$. In particular, by choosing $\hat{A}$ as the identity operator one gets $\langle \psi|\mathcal{T}\psi\rangle = 0$, i.e., any atomic state is orthogonal to the corresponding time-reversed state. Furthermore, the property $\langle \psi|\hat{A}|\mathcal{T}\psi\rangle = 0$ together with Eq. (2) imply that a generic time-reversal invariant operator is represented by a *scalar* in any

subspace of Kramers pairs: $\langle \psi_m | \hat{A} | \psi_n \rangle = \delta_{m,n} A_0$ ($m,n$=1,2) with $|\psi_1\rangle = |\psi\rangle$ and $|\psi_2\rangle = \mathcal{T}|\psi\rangle$ generic Kramers pairs and $A_0 \equiv \langle \psi | \hat{A} | \psi \rangle = \langle \mathcal{T}\psi | \hat{A} | \mathcal{T}\psi \rangle$. For example, the energy operator is represented by a scalar in a subspace of Kramers pairs.

The electric dipole moment is even under the time reversal operation ($\mathcal{T}^{-1}\hat{\mathbf{p}}\mathcal{T} = \hat{\mathbf{p}}$). Thus, from Eq. (2) one finds that $\gamma_{\mathcal{T}i,\mathcal{T}j} \equiv \langle \mathcal{T}i | \hat{\mathbf{p}} | \mathcal{T}j \rangle$ is such that $\gamma_{\mathcal{T}i,\mathcal{T}j} = \gamma_{ij}^*$. In particular, the transition dipole moments satisfy (see Fig. 1a):

$$\boldsymbol{\gamma}_d \equiv \boldsymbol{\gamma}_{g1,e1} = \boldsymbol{\gamma}_{g2,e2}^*, \qquad \boldsymbol{\gamma}_c \equiv \boldsymbol{\gamma}_{g1,e2} = -\boldsymbol{\gamma}_{g2,e1}^*. \tag{3}$$

Since any component of $\hat{\mathbf{p}}$ must be represented by a scalar in the excited and ground subspaces it is necessary that $\boldsymbol{\gamma}_{g1,g1} = \boldsymbol{\gamma}_{g2,g2}$ and $\boldsymbol{\gamma}_{g1,g2} = 0$ and that $\boldsymbol{\gamma}_{e1,e1} = \boldsymbol{\gamma}_{e2,e2}$ and $\boldsymbol{\gamma}_{e1,e2} = 0$. In this article, I suppose that the dipole moment expectation vanishes for an arbitrary stationary state ($\boldsymbol{\gamma}_{g1,g1} = \boldsymbol{\gamma}_{g2,g2} = 0$ and $\boldsymbol{\gamma}_{e1,e1} = \boldsymbol{\gamma}_{e2,e2} = 0$). In these conditions, the dipole moment matrix is completely determined by the direct transition dipole moment ($\boldsymbol{\gamma}_d$) and by the crossed transition dipole moment ($\boldsymbol{\gamma}_c$). The dipole moment operator is given by:

$$\hat{\mathbf{p}} = \hat{\mathbf{p}}^- + \hat{\mathbf{p}}^+, \tag{4a}$$

$$\hat{\mathbf{p}}^- = \boldsymbol{\gamma}_d \sigma_{g1,e1} + \boldsymbol{\gamma}_d^* \sigma_{g2,e2} + \boldsymbol{\gamma}_c \sigma_{g1,e2} - \boldsymbol{\gamma}_c^* \sigma_{g2,e1}, \tag{4b}$$

$$\hat{\mathbf{p}}^+ = \boldsymbol{\gamma}_d^* \sigma_{g1,e1}^\dagger + \boldsymbol{\gamma}_d \sigma_{g2,e2}^\dagger + \boldsymbol{\gamma}_c^* \sigma_{g1,e2}^\dagger - \boldsymbol{\gamma}_c \sigma_{g2,e1}^\dagger. \tag{4c}$$

The two elements $\boldsymbol{\gamma}_d$ and $\boldsymbol{\gamma}_c$ depend on the adopted basis of states. If $\boldsymbol{\gamma}_d$ and $\boldsymbol{\gamma}_d^*$ are linearly independent vectors, it may be shown that the crossed dipole moment can be set equal to zero after a suitable basis change *only if* $\boldsymbol{\gamma}_c$ is a linear combination of $\boldsymbol{\gamma}_d$ and $\boldsymbol{\gamma}_d^*$ in the original basis; moreover, if $\boldsymbol{\gamma}_d$ and $\boldsymbol{\gamma}_d^*$ are linearly dependent vectors, then the crossed dipole moment can set equal to zero *only if* $\boldsymbol{\gamma}_d, \boldsymbol{\gamma}_c, \boldsymbol{\gamma}_c^*$ are linearly dependent vectors in the original basis (see Appendix A). Thus, in general it is impossible to suppress the crossed coupling element $\boldsymbol{\gamma}_c$ by switching to a different eigenstate basis.

### III. Free-space electric polarizability

*A. The parametric polarizability*

In the quantum optics literature, it is usually assumed that $\boldsymbol{\gamma}_c = 0$; in such a case the transitions between the subspace of states generated by $|e_1\rangle, |g_1\rangle$ (*m*=1) and the subspace generated by $|e_2\rangle, |g_2\rangle$ (*m*=2) are forbidden. Importantly, because the spin-orbit interaction term of the atom Hamiltonian, the ground and excited states with different indices *m* can be coupled, i.e., in principle nothing forces $\boldsymbol{\gamma}_c$ to be zero. Spin-orbit interactions have been widely discussed in recent years, and, for example, it is known that they may lead to a new phase of matter, the so called topological (time-reversal invariant) insulators [3, 22, 23, 24].

Consider the interaction between a two-level system formed by two Kramers pairs and the electromagnetic field when the atom stands alone in free-space. The atomic system is prepared in an initial state of the form $|\psi_0\rangle = c_1|g_1\rangle + c_2|g_2\rangle$ with $|c_1|^2 + |c_2|^2 = 1$, i.e., in a definite ground state. For relatively weak fields and away from resonances the optical response can be linearized and the atom behaves similarly to a classical electric dipole. It is shown in the Appendix B that the (parametric) atomic polarizability satisfies:

$$\boldsymbol{\alpha}\left(\omega;|\psi_0\rangle\right) = \boldsymbol{\alpha}_R(\omega) + \boldsymbol{\alpha}_{NR}\left(\omega;|\psi_0\rangle\right), \tag{5a}$$

$$\boldsymbol{\alpha}_R = \frac{1}{\varepsilon_0\hbar}\frac{1}{2}\left(\boldsymbol{\gamma}_d \otimes \boldsymbol{\gamma}_d^* + \boldsymbol{\gamma}_d^* \otimes \boldsymbol{\gamma}_d + \boldsymbol{\gamma}_c \otimes \boldsymbol{\gamma}_c^* + \boldsymbol{\gamma}_c^* \otimes \boldsymbol{\gamma}_c\right)\left(\frac{1}{\omega_0-\omega}+\frac{1}{\omega_0+\omega}\right), \tag{5b}$$

$$\boldsymbol{\alpha}_{NR} = \frac{1}{\varepsilon_0\hbar}i\boldsymbol{\Omega}\times\mathbf{1}\left(\frac{1}{\omega_0-\omega}-\frac{1}{\omega_0+\omega}\right), \tag{5c}$$

with the *gyration vector* given by:

$$\boldsymbol{\Omega}(c_1,c_2) = -i\left(\boldsymbol{\gamma}_d\times\boldsymbol{\gamma}_c c_1^*c_2 - \boldsymbol{\gamma}_d^*\times\boldsymbol{\gamma}_c^* c_2^* c_1\right) - i\frac{1}{2}\left(|c_2|^2-|c_1|^2\right)\left(\boldsymbol{\gamma}_d\times\boldsymbol{\gamma}_d^* + \boldsymbol{\gamma}_c\times\boldsymbol{\gamma}_c^*\right). \tag{6}$$

The electric polarizability links the quantum expectation of the electric dipole moment with the local electric field as $\langle\mathbf{p}\rangle = \varepsilon_0\boldsymbol{\alpha}\cdot\mathbf{E}$. In the above, $\hbar\omega_0$ represents the energy difference between the excited and the ground states (Fig. 1a). The vector $\boldsymbol{\Omega}$ is real-valued. The tensor $\boldsymbol{\Omega}\times\mathbf{1}$ has elements $(\boldsymbol{\Omega}\times\mathbf{1})_{ij} = -\Omega_k\varepsilon_{ijk}$ with $\varepsilon_{ijk}$ the Levi-Civita symbol. The effects of spontaneous emission may be taken into account by replacing $\omega\to\omega+i\Gamma/2$ in Eq. (5), with $\Gamma$ the total spontaneous emission rate in free-space [25]. These effects are unimportant away from the resonance $\omega=\omega_0$, and for simplicity are neglected in the following analysis. Note that a dissipative $\boldsymbol{\alpha}$ due to the effects of spontaneous emission is compatible with the time reversal invariance of the system Hamiltonian. Indeed, a dissipative $\boldsymbol{\alpha}$ does not imply a non-Hermitian dynamics, but is rather a consequence of the system being open [26].

The polarizability depends explicitly on the coefficients $c_1, c_2$ that characterize the ground state, because the vector $\boldsymbol{\Omega}$ also does. The exception occurs when the direct transition dipole is linearly polarized (i.e., when the Cartesian components of $\boldsymbol{\gamma}_d$ have the same phase, so that $\boldsymbol{\gamma}_d\times\boldsymbol{\gamma}_d^* = 0$) and $\boldsymbol{\gamma}_c = 0$, which is the traditional scenario in quantum optics studies. In such a case, $\boldsymbol{\Omega}=0$ and one recovers the usual formula for the atomic polarizability (without orientation averaging) [25, 27]:

$$\boldsymbol{\alpha}^S = \frac{1}{\varepsilon_0 \hbar}\left(\frac{\boldsymbol{\gamma}_d \otimes \boldsymbol{\gamma}_d^*}{\omega_0 - \omega} + \frac{\boldsymbol{\gamma}_d^* \otimes \boldsymbol{\gamma}_d}{\omega_0 + \omega}\right). \tag{7}$$

The tensor $\boldsymbol{\Omega}\times\mathbf{1}$ in Eq. (5) is anti-symmetric, and hence it determines a gyrotropic response. When $\boldsymbol{\Omega}\neq 0$ one has $\boldsymbol{\alpha}(\omega;|\psi_0\rangle)\neq \boldsymbol{\alpha}^T(\omega;|\psi_0\rangle)$ and thereby the (parametric) optical response of the atom is generically *nonreciprocal* when it is prepared in a given definite ground state [28]. This property is consistent with the recent literature of chiral quantum optics [29]. Specifically, it has been shown both theoretically and experimentally that by preparing an elementary quantum-emitter in an initial ground-state that favors some particular circularly polarized (chiral) optical transitions it is possible to have strongly nonreciprocal light-matter interactions, e.g., highly asymmetric photon emissions, non-reciprocal absorption and modified super-radiance [29-39]. It is relevant to mention that the standard quantum optics model for a two-level atom also gives nonreciprocal responses because in general $\boldsymbol{\alpha}^S \neq \boldsymbol{\alpha}^{S,T}$ (see Eq. (7)).

An initial state of the form $|\psi_0\rangle = c_1|g_1\rangle + c_2|g_2\rangle$ is transformed by the time-reversal operation into $\mathcal{T}|\psi_0\rangle = -c_2^*|g_1\rangle + c_1^*|g_2\rangle$. It may be verified that:

$$\boldsymbol{\alpha}(\omega;|\psi_0\rangle) = \boldsymbol{\alpha}^T(\omega;\mathcal{T}|\psi_0\rangle) \tag{8}$$

where the superscript "*T*" stands for the transpose of a tensor. Thus, the polarizability tensors associated with two ground states linked by the time-reversal operation are related by matrix transposition. This is a standard property of nonreciprocal systems [28].

In the very special case $\gamma_c = 0$ with $\gamma_d$ linearly polarized, the polarizability ($\alpha = \alpha_R$) satisfies the Onsager-Casimir reciprocity relation $\alpha = \alpha^T$ [40, 41], and is independent of the initial ground state. Note that $\alpha_R$, i.e., the first component of $\alpha$ in Eq. (5a), is a symmetric tensor independent of $|\psi_0\rangle$. Furthermore, for thermal states the ground states are equiprobable and uncorrelated ($\langle |c_1|^2 \rangle_{th} = \langle |c_2|^2 \rangle_{th} = 1/2$ and $\langle c_1 c_2^* \rangle_{th} = 0$; for simplicity, it is assumed that the temperature is sufficiently small so that the probability of occupation of the excited states is negligible); thus, the expectation of the gyration vector vanishes ($\langle \Omega \rangle_{th} = 0$) and the polarizability of the two-level system reduces to $\langle \alpha \rangle_{th} = \alpha_R$, so that the optical response of thermal states is reciprocal.

### *B. Link between the gyration vector and the orbital magnetic dipole moment*

The orbital magnetic dipole moment is $\hat{\mathbf{m}} = -\gamma_e \hat{\mathcal{L}}$ where $\hat{\mathcal{L}} = \mathbf{r} \times \mathcal{P}$ is the (orbital) angular momentum operator, $\gamma_e = \dfrac{e}{2m_e} > 0$ is the (classical) gyromagnetic ratio, $m_e$ is the electron mass and $e > 0$ is the elementary charge. Here, $\mathbf{r}$ and $\mathcal{P}$ are the position and momentum operators of the electron. For a two-level system, the position operator is $\hat{\mathbf{p}}/(-e)$, with $\hat{\mathbf{p}}$ the electric dipole moment operator. In a non-relativistic approximation, the momentum can be identified with $m_e \dfrac{d\mathbf{r}}{dt} = \dfrac{m_e}{-e} \dfrac{i}{\hbar} \left[ \hat{H}_{at}, \hat{\mathbf{p}} \right]$. With the help of Eq. (4), it may be written as $\mathcal{P} = \dfrac{i\omega_0}{2\gamma_e}\left( \hat{\mathbf{p}}^- - \hat{\mathbf{p}}^+ \right)$. Hence, the orbital magnetic dipole moment operator is given by:

$$\hat{\mathbf{m}} = \dfrac{i\omega_0}{2e}\left( \hat{\mathbf{p}}^+ \times \hat{\mathbf{p}}^- - \hat{\mathbf{p}}^- \times \hat{\mathbf{p}}^+ \right). \tag{9}$$

Different from the electric dipole moment, the magnetic dipole moment is *odd* under a time-reversal ($\mathcal{T}^{-1}\hat{\mathbf{m}}\mathcal{T} = -\hat{\mathbf{m}}$). Thus, the restriction of $\hat{\mathbf{m}}$ to a subspace of Kramers pairs does not need to be represented by a scalar. In particular, the expectation of $\hat{\mathbf{m}}$ is generically ground state dependent. Straightforward calculations show that $\langle\hat{\mathbf{m}}\rangle_0 \equiv \langle\psi_0|\hat{\mathbf{m}}|\psi_0\rangle$ is given by:

$$\langle\hat{\mathbf{m}}\rangle_0 = -\frac{\omega_0}{e}\mathbf{\Omega}. \qquad (10)$$

Thus, the gyration vector $\mathbf{\Omega}$ determines the expectation of the ground-state orbital magnetic dipole moment. In simple terms, due to the spin-orbit coupling the atom behaves as a tiny magnet whose magnetization depends on the ground state. Ground states linked by the time-reversal operator have anti-parallel (additive symmetric) orbital magnetic dipole moments. Indeed, because $\hat{\mathbf{m}}$ is odd under a time-reversal it is necessary that $\langle g_1|\hat{\mathbf{m}}|g_1\rangle = -\langle g_2|\hat{\mathbf{m}}|g_2\rangle$.

## IV. Casimir physics

So far, it was assumed that the atom is far from other material bodies. The rest of the article focuses on the interaction of the Kramers two-level atom with the electromagnetic vacuum. The quantum emitter is placed at a distance *d* from a macroscopic planar material interface, as depicted in Fig. 1b.

### A. *Casimir interaction energy*

The Casimir interaction energy, also referred to as the van der Waals (vdW) potential, gives the energy shift of a given atomic ground state due to the interactions with the vacuum of the quantized electromagnetic field [25]. To lowest order perturbation theory and for a generic atom "ground state" $|\psi_0\rangle$, it can be written in terms of the polarizability as follows [25, 42]:

$$\mathcal{E}_{\text{int}}(|\psi_0\rangle) = \frac{-\hbar}{2\pi}\int_0^\infty \text{tr}\{\boldsymbol{\alpha}(i\xi;|\psi_0\rangle)\cdot\mathbf{C}^{\text{int}}(i\xi)\}d\xi, \quad (11)$$

with $\text{tr}\{..\}$ the trace operator and $\mathbf{C}^{\text{int}}$ an interaction tensor that describes the coupling of the two-level system and the material substrate (see Appendix C). The polarizability is evaluated along the imaginary frequency axis ($\omega = i\xi$) where the material response does not exhibit any resonant features [43]. For now, it is assumed that $k_B T \ll \hbar\omega_0$ and $d \ll \lambda_T$ with $\lambda_T = hc/k_B T$ the thermal wavelength, so that the temperature corrections are negligible.

Evidently, $\mathcal{E}_{\text{int}}$ may depend on $|\psi_0\rangle$ because $\boldsymbol{\alpha}$ also does. Therefore, initial states of the form $|\psi_0\rangle = c_1|g_1\rangle + c_2|g_2\rangle$ may suffer different energy shifts. This implies that different from the free-atom case, when the atom is in the vicinity of a material surface there may be an energy cost to push it from a given free-atom ground state to another ground state.

For future reference, it is noted that in the standard model of quantum optics, i.e., when the two level atom has uniquely 2 eigenstates so that the polarizability is given by Eq. (7), the Casimir interaction energy is given by:

$$\mathcal{E}_{\text{int}}^S(\boldsymbol{\gamma}_d) = \frac{-1}{2\pi\varepsilon_0}\int_0^\infty \left(\frac{\boldsymbol{\gamma}_d^*\cdot\mathbf{C}^{\text{int}}\cdot\boldsymbol{\gamma}_d}{\omega_0 - \omega} + \frac{\boldsymbol{\gamma}_d\cdot\mathbf{C}^{\text{int}}\cdot\boldsymbol{\gamma}_d^*}{\omega_0 + \omega}\right)d\xi. \quad (12)$$

The superscript "S" refers to the standard model approximation.

## B. Reciprocal substrate

From Eq. (5) the interaction energy can be written as $\mathcal{E}_{int} = \frac{-\hbar}{4\pi} \int_0^\infty \text{tr}\{(\boldsymbol{\alpha}_R + \boldsymbol{\alpha}_{NR}) \cdot \mathbf{C}^{int}\} d\xi$. For a reciprocal substrate, e.g., a standard metal surface, the interaction tensor $\mathbf{C}^{int}$ is symmetric. The trace of the product of symmetric and anti-symmetric matrices vanishes. Using this property one readily finds that:

$$\begin{aligned}\mathcal{E}_{int} &= \frac{-\hbar}{2\pi} \int_0^\infty \text{tr}\{\boldsymbol{\alpha}_R \cdot \mathbf{C}^{int}\} d\xi \\ &= \frac{-1}{2\pi\varepsilon_0} \int_0^\infty \left(\frac{1}{\omega_0 - \omega} + \frac{1}{\omega_0 + \omega}\right)\left(\boldsymbol{\gamma}_d^* \cdot \mathbf{C}^{int} \cdot \boldsymbol{\gamma}_d + \boldsymbol{\gamma}_c^* \cdot \mathbf{C}^{int} \cdot \boldsymbol{\gamma}_c\right) d\xi\end{aligned} \quad (13)$$

Hence, the interaction energy (i.e., the energy shift suffered by $|\psi_0\rangle$) is independent of the considered ground state $|\psi_0\rangle$. In particular, the energy shift is insensitive to the nonreciprocity of the (parametric) polarizability response.

For a reciprocal substrate ($\mathbf{C}^{int}$ is symmetric), the interaction energy calculated with the standard two-level atom model [Eq. (12)] satisfies $\mathcal{E}_{int}^S = \mathcal{E}_{int}|_{\gamma_c=0}$, with $\mathcal{E}_{int}|_{\gamma_c=0}$ the interaction energy calculated for a Kramers two-level system [Eq. (13)] with no spin-orbit coupling ($\gamma_c = 0$). Thus, if there is no spin-orbit coupling and if the substrate is reciprocal, it makes no difference to compute the Casimir energy (ground state energy shift of the atom) using the standard two-level atom model or the Kramers pairs model: the result is exactly the same.

Consider now that the substrate is a metal half-space described by a Drude-type dispersion model $\varepsilon(\omega) = 1 - 2\omega_{sp}^2 / \omega(\omega + i\Gamma_m)$ with $\omega = \omega_{sp}$ the surface plasmon (SPP) resonance and $\Gamma_m$ a damping factor. Using a quasi-static approximation, one finds the following explicit formula for the interaction dyadic (see Appendix D):

$$\mathbf{C}_{int}(\omega) \approx \frac{1}{32\pi d^3} \frac{\varepsilon(\omega)-1}{\varepsilon(\omega)+1} (\hat{\mathbf{x}} \otimes \hat{\mathbf{x}} + \hat{\mathbf{y}} \otimes \hat{\mathbf{y}} + 2\hat{\mathbf{z}} \otimes \hat{\mathbf{z}}). \quad (14)$$

As shown in Fig. 2, this quasi-static approximation gives results nearly indistinguishable from the exact calculation based on Eq. (C4) of Appendix C for $\omega = i\xi$ in the imaginary frequency axis and when the atom-metal distance is deeply subwavelength ($d\omega_{sp}/c \ll 1$). Note that in the imaginary frequency axis $\mathbf{C}_{int}$ is real valued and exhibits a monotonic decreasing behavior with no resonant features. Due to this reason, $\mathbf{C}_{int}$ is little affected by realistic metal loss ($\Gamma_m = 0.1\omega_{sp}$) in the imaginary frequency axis, and thereby the interaction Casimir energy has the same property.

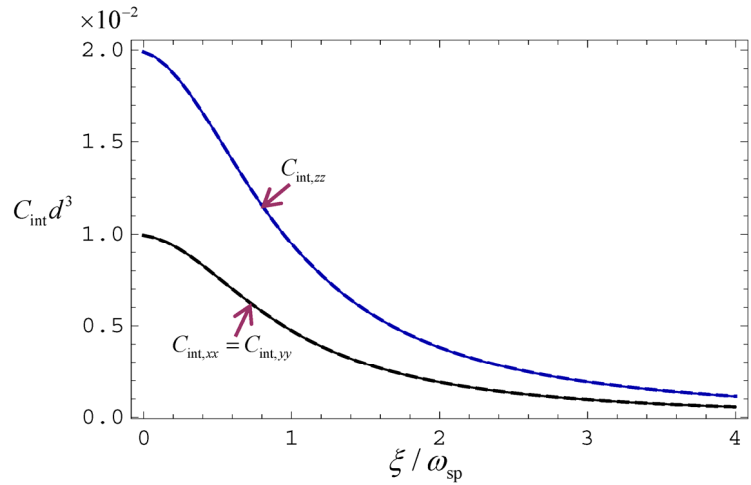

**Fig. 2** Interaction tensor evaluated for imaginary frequencies ($\omega = i\xi$) for $d = 0.01c/\omega_{sp}$ and $\Gamma_m = 0.1\omega_{sp}$. Solid lines: Exact result [Eq. (C4)]. Dashed lines: Quasi-static approximation [Eq. (14)]. The plot is nearly unchanged for smaller values of the distance $d$.

In the lossless limit ($\Gamma_m \to 0^+$), one has $\dfrac{\varepsilon(\omega)-1}{\varepsilon(\omega)+1} \approx \dfrac{\omega_{sp}^2}{\omega_{sp}^2 - \omega^2}$. Using the auxiliary identity

$$\frac{1}{2\pi}\int_0^\infty \left(\frac{1}{\omega_0-\omega}+\frac{1}{\omega_0+\omega}\right)\frac{\omega_{sp}^2}{\omega_{sp}^2-\omega^2}\bigg|_{\omega=i\xi} d\xi = \frac{\omega_{sp}/2}{\omega_{sp}+\omega_0}, \quad (15)$$

and Eqs. (13)-(14) and one obtains the following closed analytical formula for the Casimir energy:

$$\mathcal{E}_{int} = \frac{-1}{64\pi\varepsilon_0 d^3}\frac{\omega_{sp}}{\omega_{sp}+\omega_0}\left(\boldsymbol{\gamma}_d^*\cdot\tilde{\mathbf{C}}\cdot\boldsymbol{\gamma}_d + \boldsymbol{\gamma}_c^*\cdot\tilde{\mathbf{C}}\cdot\boldsymbol{\gamma}_c\right) \quad (16)$$

with $\tilde{\mathbf{C}} = \hat{\mathbf{x}}\otimes\hat{\mathbf{x}} + \hat{\mathbf{y}}\otimes\hat{\mathbf{y}} + 2\hat{\mathbf{z}}\otimes\hat{\mathbf{z}}$.

## C. Nonreciprocal substrate

Suppose now that the substrate is nonreciprocal so that $\mathbf{C}^{int}$ does not need to be a symmetric tensor. For simplicity, it is assumed that the crossed transition dipole moment vanishes so that $\boldsymbol{\gamma}_c = 0$. Using Eq. (B7), the interaction energy for the Kramers pairs model is:

$$\begin{aligned}\mathcal{E}_{int}\big|_{\boldsymbol{\gamma}_c=0} &= \frac{-1}{2\pi\varepsilon_0}|c_1|^2\int_0^\infty\left(\frac{\boldsymbol{\gamma}_d^*\cdot\mathbf{C}^{int}\cdot\boldsymbol{\gamma}_d}{\omega_0-\omega}+\frac{\boldsymbol{\gamma}_d\cdot\mathbf{C}^{int}\cdot\boldsymbol{\gamma}_d^*}{\omega_0+\omega}\right)\bigg|_{\omega=i\xi} d\xi \\ &+ \frac{-1}{2\pi\varepsilon_0}|c_2|^2\int_0^\infty\left(\frac{\boldsymbol{\gamma}_d\cdot\mathbf{C}^{int}\cdot\boldsymbol{\gamma}_d^*}{\omega_0-\omega}+\frac{\boldsymbol{\gamma}_d^*\cdot\mathbf{C}^{int}\cdot\boldsymbol{\gamma}_d}{\omega_0+\omega}\right)\bigg|_{\omega=i\xi} d\xi.\end{aligned} \quad (17)$$

Remarkably, for a nonreciprocal substrate the interaction energy *generally depends on* $|\psi_0\rangle$. The interaction energy can be written in terms of $\mathcal{E}_{int}^S(\boldsymbol{\gamma}_d)$ [Eq. (12)] as follows:

$$\mathcal{E}_{int}\big|_{\boldsymbol{\gamma}_c=0} = |c_1|^2\mathcal{E}_{int}^S(\boldsymbol{\gamma}_d) + |c_2|^2\mathcal{E}_{int}^S(\boldsymbol{\gamma}_d^*). \quad (18)$$

Thus, in the Kramers pairs model, the Casimir energy is a weighted sum of the Casimir energies associated with the individual (uncoupled) two-level atom components.

Furthermore, in Appendix E it shown that the ground-state physics can be described by an effective Hamiltonian that takes into account the interaction with the quantized electromagnetic field. The effective Hamiltonian $H_{ef} = [h_{m,n}]$ ($m,n = 1,2$) acts on the ground-state coordinates $(c_1 \ c_2)^T$ with $h_{m,n}$ determined by Eq. (E9). In the reciprocal case $H_{ef}$ reduces to a scalar, as the ground state is necessarily degenerate. In contrast, for a nonreciprocal substrate $H_{ef}$ is a nontrivial 2×2 matrix. Thereby, the interactions with the material substrate can lift the ground state degeneracy. When $\gamma_c = 0$ the energy eigenstates of the interacting system are $|g_1\rangle$ and $|g_2\rangle$. Typically the two eigenstates are associated with different energies, consistent with Eq. (18). Even though a generic state of the form $|\psi_0\rangle = c_1 |g_1\rangle + c_2 |g_2\rangle$ is *not* a stationary state of the interacting system, the energy expectation is always determined by the Casimir energy $\langle \psi_0 | H_{ef} | \psi_0 \rangle = \mathcal{E}_{int}(|\psi_0\rangle)$ (see Appendix E).

To illustrate the ideas, suppose that the substrate is an electron gas (metal) biased with a static magnetic field oriented along the *y*-direction, i.e., parallel to the interface with air (*z*=0). The nonreciprocal material is characterized by the plasma frequency $\omega_p$ and by the cyclotron frequency $\omega_c$. The surface plasmon resonance is $\omega_{sp} = \omega_p / \sqrt{2}$. The dispersive model of the gyrotropic permittivity tensor is the same as in Ref. [42]. The fluctuation-induced forces in this material platform were characterized in Ref. [42] using a quasi-static approximation for the standard two-level system model. From Eq. (44) of Ref. [42], it follows that $\mathcal{E}_{int}^S$ can be written as:

$$\mathcal{E}_{int}^S(\boldsymbol{\gamma}_d) = \frac{1}{16\pi\varepsilon_0 d^3} \frac{-1}{2\pi} \int_0^{2\pi} \frac{a_\theta \omega_\theta}{\omega_0 + \omega_\theta} \left| (i\cos\theta \hat{\mathbf{x}} + i\sin\theta \hat{\mathbf{y}} + \hat{\mathbf{z}}) \cdot \boldsymbol{\gamma}_d^* \right|^2 d\theta, \qquad (19)$$

where $a_\theta, \omega_\theta$ are functions defined in Ref. [42] and which are independent of $\gamma_d$. The Casimir interaction energy for the Kramers pairs model can be evaluated using the above formula and Eq. (18). As an aside, I note that when there is no magnetic bias ($\omega_c = 0$), one has $a_\theta = 1/2$ and $\omega_\theta = \omega_{sp}$, so that Eq. (19) reduces to the right-hand side of Eq. (16) with $\gamma_c = 0$.

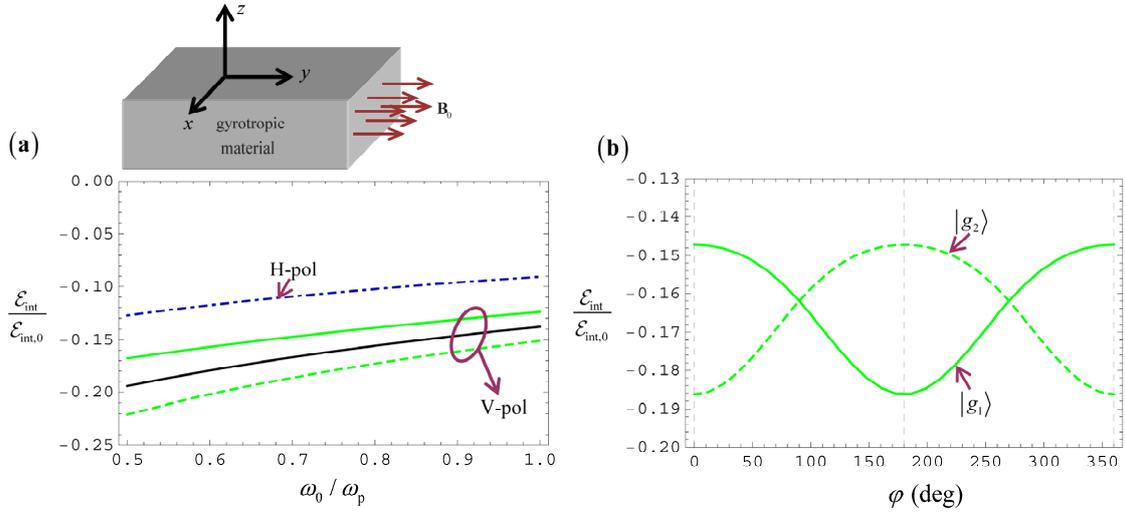

**Fig. 3** Normalized interaction energy for a two-level system formed by two Kramers pairs placed nearby a gyrotropic magnetized plasma with $\omega_c = 0.5\omega_p$. **a)** $\mathcal{E}_{int}$ as a function of the atomic transition frequency for i) $\boldsymbol{\gamma}_d = \gamma_d(\hat{\mathbf{x}} \pm i\hat{\mathbf{y}})/\sqrt{2}$ and an arbitrary ground state (H-pol) ii) $\boldsymbol{\gamma}_d = \gamma_d(\hat{\mathbf{x}} - i\hat{\mathbf{z}})/\sqrt{2}$ (V-pol) and for the states $|g_1\rangle$ (solid green line), $(|g_1\rangle + e^{i\delta}|g_2\rangle)/\sqrt{2}$ (solid black line) and $|g_2\rangle$ (dashed green line). **b)** $\mathcal{E}_{int}$ as a function of the direction $\varphi$ of the atom polarization plane for the stationary states of the interacting system and $\omega_0 = 0.7\omega_p$.

Figure 3a shows the normalized Casimir energy as a function of the two-level atom transition frequency $\omega_0$ for different free-atom ground states. The energy normalization factor is $\mathcal{E}_{int,0} = |\boldsymbol{\gamma}_d|^2 / (16\pi\varepsilon_0 d^3)$. The direct transition dipole moment is circularly polarized. For a vertical plane of polarization (V-pol) it is of the form $\boldsymbol{\gamma}_d = \gamma_d(\hat{\mathbf{x}} - i\hat{\mathbf{z}})/\sqrt{2}$, whereas for an horizontal plane of polarization (H-pol) it is given by $\boldsymbol{\gamma}_d = \gamma_d(\hat{\mathbf{x}} \pm i\hat{\mathbf{y}})/\sqrt{2}$. The Casimir force acting on the atom is directed along *z*. For stationary states, it is related to the Casimir energy as $\mathcal{F}_{C,z} = -\partial \mathcal{E}_{int}/\partial d = 3\mathcal{E}_{int}/d < 0$ (attractive force) and has a variation with $\omega_0$ analogous to the variation of $\mathcal{E}_{int}$ in normalized unities (not shown).

The atomic energy shift $\mathcal{E}_{int}$ is ground-state independent for H-pol. In this case, $H_{ef}$ is a scalar and the ground-state of the interacting system is degenerate. In contrast, for V-pol $\mathcal{E}_{int}$ varies with $|\psi_0\rangle$. In particular, the Kramers pairs $|g_1\rangle$ and $|g_2\rangle$ suffer different energy shifts when the polarization plane is vertical. This energy level splitting is due to the nonreciprocal magnetic bias and is reminiscent of (but not the same as) the Zeeman effect. Note that the present analysis neglects the coupling of the atomic spin with the static magnetic bias, which is mechanism responsible for the Zeeman effect.

The Casimir torque may act to reorient the atom plane of polarization [11, 45, 46, 47]. To study this effect, I show in Fig. 3b how the interaction energy varies with the orientation $\varphi$ of the vertical plane of polarization. The corresponding transition dipole moment is $\boldsymbol{\gamma}_d(\varphi) = \gamma_d(\hat{\boldsymbol{\rho}} - i\hat{\mathbf{z}})/\sqrt{2}$, with $\hat{\boldsymbol{\rho}} = \cos\varphi\hat{\mathbf{x}} + \sin\varphi\hat{\mathbf{y}}$ and $\varphi$ the angle of the polarization plane with respect to the *x*-axis. For stationary states, the Casimir torque is $\boldsymbol{\tau}_C = -\partial_\varphi \mathcal{E}_{int} \hat{\mathbf{z}}$.

As seen in Fig. 3b, the energy minimum occurs at $\varphi = 0°$ when $|\psi_0\rangle = |g_2\rangle$ and at $\varphi = 180°$ when $|\psi_0\rangle = |g_1\rangle$, which thereby are the ground-state dependent orientations for

a stable equilibrium. Thus, the Casimir torque acts to reorient the atom plane of polarization towards the direction $\varphi = 0°$ ($\varphi = 180°$) when the initial ground state is $|g_2\rangle$ ($|g_1\rangle$). In both cases, the plane of polarization is perpendicular to the bias magnetic field (see also Ref. [11]). Interestingly, the atomic polarizability at the equilibrium configurations is identical in both cases ($\mathbf{\alpha}(\omega;|g_2\rangle_{\varphi=0°}) = \mathbf{\alpha}(\omega;|g_1\rangle_{\varphi=180°})$). Thus, the *equilibrium* state of the atom has properties that are independent of the initial ($|g_1\rangle$ or $|g_2\rangle$) ground state. Note that it is implicit that the atom is free to rotate under the action of the Casimir torque. Furthermore, the gyration vector is the same for both $|g_1\rangle_{\varphi=180°}, |g_2\rangle_{\varphi=0°}$ ($\mathbf{\Omega} = -\hat{\mathbf{y}}|\gamma_d|^2/2$), i.e., it is anti-parallel to the bias magnetic field. Consequently, from Eq. (10) the magnetic dipole moment expectation $\langle\hat{\mathbf{m}}\rangle$ of the stable ground-state of the interacting system is directed along $+\hat{\mathbf{y}}$. The tiny magnetic field induced by $\langle\hat{\mathbf{m}}\rangle$ acts to demagnetize the substrate (note that the atom lies above the substrate).

In summary, for a nonreciprocal substrate the ground-state does not need to be degenerate. When $\gamma_c$ vanishes, the stationary states of the interacting system are $|g_1\rangle$ and $|g_2\rangle$. For a circularly polarized atom, the Casimir torque acts to reorient the polarization plane in such a way that it is vertical and perpendicular to the magnetic bias.

## V. Spontaneous symmetry breaking of rotational symmetry

Next, I analyze the equilibrium positions of a Kramers two-level atom when it is placed nearby a reciprocal metal surface. It is supposed that the atom has a rotational symmetry about some reference axis. It is shown that the ground state of the combined system (atom + substrate) may have a broken symmetry depending on the relative strength of dipole transitions with linear and circular polarization.

## A. Casimir torque

It is useful to obtain a general formula for the Casimir torque acting on the atomic system. As is well-known, the Casimir torque along a generic direction $\hat{\mathbf{n}}$ can be found from the variation of the Casimir energy under a rotation about $\hat{\mathbf{n}}$ [11, 44, 45]. The transition dipole moment matrix elements transform as $\boldsymbol{\gamma} \to \mathbf{R}_\alpha \cdot \boldsymbol{\gamma}$ under a rotation by an angle $\alpha$. Here,

$$R_\alpha(\hat{\mathbf{n}}) = e^{\alpha \hat{\mathbf{n}} \times \mathbf{1}} = \hat{\mathbf{n}} \otimes \hat{\mathbf{n}} + (\mathbf{1} - \hat{\mathbf{n}} \otimes \hat{\mathbf{n}})\cos\alpha + \sin\alpha(\hat{\mathbf{n}} \times \mathbf{1}) \qquad (20)$$

is the rotation matrix. The interaction energy may be regarded a function of the transition dipole moment vector amplitudes $\mathcal{E}_{\text{int}} = \mathcal{E}_{\text{int}}(\boldsymbol{\gamma}_d, \boldsymbol{\gamma}_d^*, \boldsymbol{\gamma}_c, \boldsymbol{\gamma}_c^*)$, so that under a rotation of $\alpha$ around the direction $\hat{\mathbf{n}}$ one has $\mathcal{E}_{\text{int}}(\alpha) = \mathcal{E}_{\text{int}}(e^{\alpha\hat{\mathbf{n}}\times\mathbf{1}} \cdot \boldsymbol{\gamma}_d, e^{\alpha\hat{\mathbf{n}}\times\mathbf{1}} \cdot \boldsymbol{\gamma}_d^*, e^{\alpha\hat{\mathbf{n}}\times\mathbf{1}} \cdot \boldsymbol{\gamma}_c, e^{\alpha\hat{\mathbf{n}}\times\mathbf{1}} \cdot \boldsymbol{\gamma}_c^*)$. The Casimir torque $\boldsymbol{\tau}_C$ satisfies $\hat{\mathbf{n}} \cdot \boldsymbol{\tau}_C = -\partial_\alpha \mathcal{E}_{\text{int}}|_{\alpha=0}$, which yields:

$$\hat{\mathbf{n}} \cdot \boldsymbol{\tau}_C = -\sum_{\boldsymbol{\gamma}=\boldsymbol{\gamma}_d, \boldsymbol{\gamma}_d^*, \boldsymbol{\gamma}_c, \boldsymbol{\gamma}_c^*} (\hat{\mathbf{n}} \times \boldsymbol{\gamma}) \cdot \partial_\gamma \mathcal{E}_{\text{int}}, \qquad (21)$$

with $\partial_\gamma = \partial/\partial\boldsymbol{\gamma}$. Since $\hat{\mathbf{n}}$ is arbitrary the torque can be expressed through the compact formula:

$$\boldsymbol{\tau}_C = -\sum_{\boldsymbol{\gamma}=\boldsymbol{\gamma}_d, \boldsymbol{\gamma}_d^*, \boldsymbol{\gamma}_c, \boldsymbol{\gamma}_c^*} \boldsymbol{\gamma} \times \partial_\gamma \mathcal{E}_{\text{int}}. \qquad (22)$$

## B. Metal surface

Applying Eq. (22) to a (reciprocal) metal surface with interaction energy given by Eq. (16), one gets:

$$\boldsymbol{\tau}_{\rm C} = \frac{1}{64\pi\varepsilon_0 d^3} \frac{\omega_{\rm sp}}{\omega_{\rm sp}+\omega_0} \left( \boldsymbol{\gamma}_{\rm d}^* \times (\tilde{\bf C}\cdot \boldsymbol{\gamma}_{\rm d}) + \boldsymbol{\gamma}_{\rm d} \times (\tilde{\bf C}\cdot \boldsymbol{\gamma}_{\rm d}^*) + \boldsymbol{\gamma}_{\rm c}^* \times (\tilde{\bf C}\cdot \boldsymbol{\gamma}_{\rm c}) + \boldsymbol{\gamma}_{\rm c} \times (\tilde{\bf C}\cdot \boldsymbol{\gamma}_{\rm c}^*) \right). \qquad (23)$$

I used the fact that $\tilde{\bf C}$ is a symmetric tensor. Noting that $\tilde{\bf C} = {\bf 1} + \hat{\bf z}\otimes\hat{\bf z}$ the above result can be rewritten as:

$$\boldsymbol{\tau}_{\rm C} = \frac{1}{32\pi\varepsilon_0 d^3} \frac{\omega_{\rm sp}}{\omega_{\rm sp}+\omega_0} {\rm Re}\left\{ (\boldsymbol{\gamma}_{\rm d}^*\times\hat{\bf z})(\hat{\bf z}\cdot\boldsymbol{\gamma}_{\rm d}) + (\boldsymbol{\gamma}_{\rm c}^*\times\hat{\bf z})(\hat{\bf z}\cdot\boldsymbol{\gamma}_{\rm c})\right\}. \qquad (24)$$

In the rest of the article, I focus on an atomic system that is invariant under arbitrary rotations about some reference axis. When the reference axis is aligned with the $z$-direction, the entire system (atom + metallic substrate) is invariant under arbitrary rotations around $z$.

In order that the Kramers two-level atom is invariant under arbitrary rotations around $z$ it is necessary that both $\boldsymbol{\gamma}_{\rm d}$ and $\boldsymbol{\gamma}_{\rm c}$ are eigenvectors of $R_\alpha(\hat{\bf z})$. Thus, the direct and crossed dipole moments must proportional to either $\hat{\bf z}$ or ${\bf e}_+ = (\hat{\bf x}+i\hat{\bf y})/\sqrt{2}$ or ${\bf e}_- = (\hat{\bf x}-i\hat{\bf y})/\sqrt{2}$. The crossed dipole term cannot be eliminated with a suitable change of basis when the two dipole moments have a different polarization type; specifically one of the dipole moments must be linearly polarized and the other one must be circularly polarized. In the following, I consider the case

$$\boldsymbol{\gamma}_{\rm d}\big|_{\text{axis align. }z} = \gamma_{\rm d}\hat{\bf z}, \qquad \boldsymbol{\gamma}_{\rm c}\big|_{\text{axis align. }z} = \gamma_{\rm c}(\hat{\bf x}-i\hat{\bf y})\frac{1}{\sqrt{2}}. \qquad (25)$$

Since $\boldsymbol{\gamma}_{\rm d}, \boldsymbol{\gamma}_{\rm c}, \boldsymbol{\gamma}_{\rm c}^*$ are linearly independent it is impossible to eliminate the crossed dipole element with a basis change. Evidently, when the axis of symmetry of the atom is aligned along the generic radial direction $\hat{\bf r}$, Eq. (25) must be replaced by:

$$\boldsymbol{\gamma}_d(\theta,\varphi) = \gamma_d \hat{\mathbf{r}}, \qquad \boldsymbol{\gamma}_c(\theta,\varphi) = \gamma_c\left(\hat{\boldsymbol{\theta}} - i\hat{\boldsymbol{\varphi}}\right)\frac{1}{\sqrt{2}}. \qquad (26)$$

Here, $\hat{\mathbf{r}}, \hat{\boldsymbol{\theta}}, \hat{\boldsymbol{\varphi}}$ are unit vectors associated with a spherical coordinate system attached to the Cartesian reference frame. Substituting the above formulas into Eqs. (16) and (24) it is found that the Casimir energy and torque are given by:

$$\mathcal{E}_{int} = \frac{-1}{64\pi\varepsilon_0 d^3}\frac{\omega_{sp}}{\omega_{sp}+\omega_0}\left(2|\gamma_d|^2 + |\gamma_c|^2 + \left(-|\gamma_d|^2 + \frac{|\gamma_c|^2}{2}\right)\sin^2\theta\right), \qquad (27)$$

$$\boldsymbol{\tau}_C = \frac{1}{64\pi\varepsilon_0 d^3}\frac{\omega_{sp}}{\omega_{sp}+\omega_0}\sin 2\theta\left(-|\gamma_d|^2 + \frac{|\gamma_c|^2}{2}\right)\hat{\boldsymbol{\varphi}}. \qquad (28)$$

Clearly, the torque acts to rotate the symmetry axis of the Kramers two-level atom in a vertical plane $\varphi = const$. The torque vanishes when the atom symmetry axis is oriented along $\theta = 0º, 180º$, i.e., when the atom and the substrate symmetry axes are aligned, or when $\theta = 90º$, i.e., when the two symmetry axes are perpendicular.

## C. Symmetry breaking

The point of stable equilibrium determined by the Casimir torque depends on the relative strength of the direct and crossed dipole moments. If $|\gamma_d| > |\gamma_c|/\sqrt{2}$, i.e., when the strength of linearly polarized (direct) dipole transitions dominates, the positions of stable equilibrium are $\theta = 0º, 180º$, so that the atom symmetry axis is aligned with the $z$-direction and the equilibrium ground state has the same rotational symmetry as the system (Fig. 4a, top). In particular, the minimum of $\mathcal{E}_{int}$ occurs for $\theta = 0º$ (Fig. 4a, bottom). Note that in the lower panels of Fig. 4 the angle $\theta$ is measured along the radial direction.

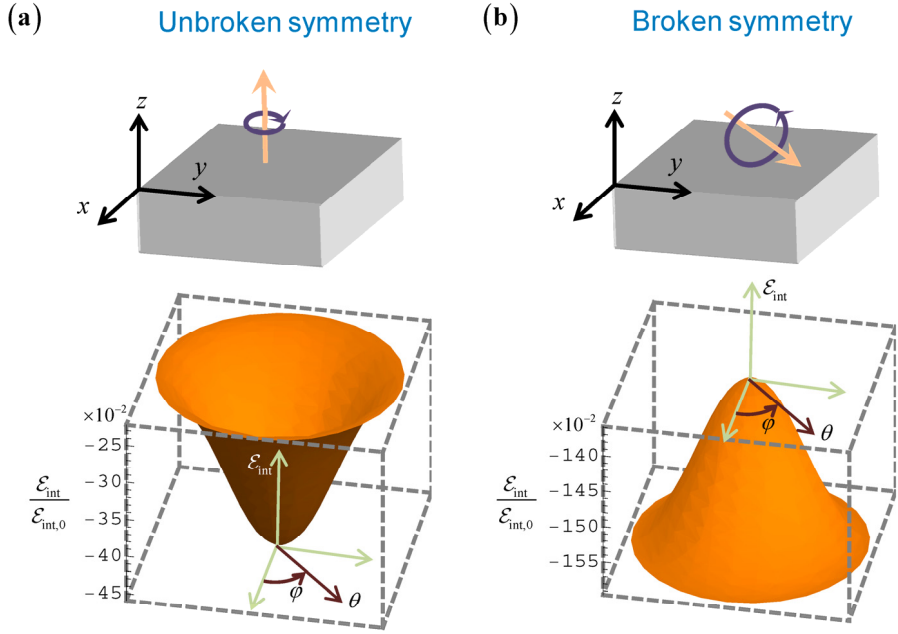

**Fig. 4** Normalized Casimir energy as a function of the orientation of the symmetry axis of a Kramers two-level atom with $\omega_0 = 0.1\omega_{sp}$. The atom symmetry axis orientation is determined by the spherical coordinates $\theta$ and $\varphi$. In the lower panels, the angle $\theta$ is represented as a radial coordinate in the interval $0° < \theta < 90°$. The Casimir torque acts to push the system state to the valley region. **a)** $\gamma_c = 0$ (linear polarization dominates), corresponding to an unbroken rotation symmetry. The atom symmetry axis is vertical with respect to the interface (upper panel). **b)** $\gamma_c = 2\gamma_d$ (circular polarization dominates), corresponding to a spontaneously broken rotation symmetry. The atom symmetry axis is horizontal with respect to the interface (upper panel).

Remarkably, when $|\gamma_d| < |\gamma_c|/\sqrt{2}$ and the strength of the chiral (circularly polarized) transitions becomes dominant, the position of stable equilibrium corresponds to $\theta = 90°$ and arbitrary $\varphi$, which are the minima of $\mathcal{E}_{int}$ (Fig. 4b, bottom). The configurations of stable equilibrium form a continuum. For any of the equilibrium points the atom symmetry axis is horizontal with respect to the substrate (Fig. 4b, top). Any two equilibrium configurations differ by some rotation with respect the *z*-direction; a given equilibrium configuration does not exhibit any particular symmetry. Evidently, the equilibrium ground state has less symmetry than the system itself: the rotation symmetry is spontaneously broken by the vacuum fluctuations. The point $|\gamma_d| = |\gamma_c|/\sqrt{2}$ marks a phase transition between the preserved and the spontaneously broken rotational symmetries. The spontaneous symmetry breaking due to the interaction of an atom with a metallic surface is reminiscent of other forms of (parity) symmetry breaking that occur in natural compounds and organic molecules (e.g., in polar molecules such as ammonia or in sugar molecules) [17].

A simple mechanical analogue of the system consists of a cylindrical bottle standing in still water. If the bottle stands vertical with respect to the water surface the system has rotational symmetry. However, this equilibrium point is unstable: any tiny perturbation of the water surface will make the bottle fall along some direction; the stable equilibrium position corresponds to a situation for which the bottle symmetry axis is horizontal with the respect to the interface. The rotational symmetry of the system is spontaneously broken by fluctuations on the water surface.

It is relevant to mention that the continuous variation between the state with $\gamma_c = 0$, which has a ground with a preserved rotational symmetry, and the state with $\gamma_d = 0$, which has a spontaneously broken rotational symmetry, is only possible using the Kramers pairs two-level atom model introduced here. Indeed, in the standard two-level atom model it is impossible (without suppressing the light-matter interactions) to deform continuously $\gamma_{d1}\hat{\mathbf{z}}$ into $\gamma_{d2}(\hat{\mathbf{x}} - i\hat{\mathbf{y}})\frac{1}{\sqrt{2}}$ preserving the rotational symmetry of the atom along the $z$-axis; note that a transition dipole moment of the form $\gamma_{d1}\hat{\mathbf{z}} + \gamma_{d2}(\hat{\mathbf{x}} - i\hat{\mathbf{y}})\frac{1}{\sqrt{2}}$ does not describe an atom invariant under rotations around the z-axis because it is not an eigenvector of $R_\alpha(\hat{\mathbf{z}})$. It is also pertinent to point out that the roles of $\boldsymbol{\gamma}_d$ and $\boldsymbol{\gamma}_c$ can be exchanged. Indeed, under a change of basis $|e_1\rangle \to |e_2\rangle$ and $|e_2\rangle \to -|e_1\rangle$ the meaning of the vectors is swapped (see Appendix A). Thus, the general requirement for the spontaneous symmetry breaking is that the chiral-type transitions predominate. This means that the platforms that are currently being studied in the context of chiral quantum optics are potentially suitable for the observation of the effect [29]. I also note that the spontaneous symmetry breaking can be predicted with the standard two-level atom model using $\boldsymbol{\gamma}_d(\theta,\varphi) = \gamma_d(\hat{\boldsymbol{\theta}} - i\hat{\boldsymbol{\varphi}})/\sqrt{2}$. However, in the standard model the atomic response has a broken time-reversal symmetry, different from the Kramers pairs model which respects the time-reversal invariance.

Figure 5a represents the Casimir torque as a function of the orientation of the symmetry axis $\theta$. As seen, the torque's sign is linked to the sign of $|\gamma_c| - \sqrt{2}|\gamma_d|$. When $|\gamma_c| > \sqrt{2}|\gamma_d|$ (blue lines) the torque acts to reorient the symmetry axis so that it becomes parallel to the substrate (direction $\theta = 90º$); otherwise, it acts to reorient the symmetry axis along the z-axis ($\theta = 0º, 180º$). The torque vanishes when $|\gamma_c| = \sqrt{2}|\gamma_d|$ which is the point of the phase transition.

So far, the analysis neglected the effect of temperature. Figures 5bi and 5bii show the Casimir (free) energy as a function of $\theta$ for configurations with an unbroken and with a broken rotational symmetry, respectively, and different values of the temperature. The free Casimir energy is determined in a standard way by replacing the integration over the imaginary frequency axis in Eq. (11) by a summation over Matsubara frequencies ($\xi_l = \omega_T l$ with $\omega_T = 2\pi k_B T / \hbar$ and $l=0,1,2,\ldots$) [48-50]:

$$\mathcal{E}_{\text{int}} = -k_B T \sum_{l=0,1,2,\ldots} \left(1 - \frac{1}{2}\delta_{l,0}\right) \text{tr}\left\{\boldsymbol{\alpha}(i\xi_l) \cdot \mathbf{C}^{\text{int}}(i\xi_l)\right\}. \tag{29}$$

For $\omega_T < \omega_0$, the temperature corrections are insignificant (see the solid and dashed black lines in Fig. 5b). For typical values of the atomic transition frequency, the condition $\omega_T = \omega_0$ corresponds to temperatures on the order of a few thousand of Kelvin. For even higher temperatures, the term $l=0$ in Eq. (29) becomes dominant and the free energy becomes roughly proportional to the temperature. The results of Fig. 5b suggest that the phenomenon of spontaneous symmetry breaking may be very robust to the effects of a finite temperature.

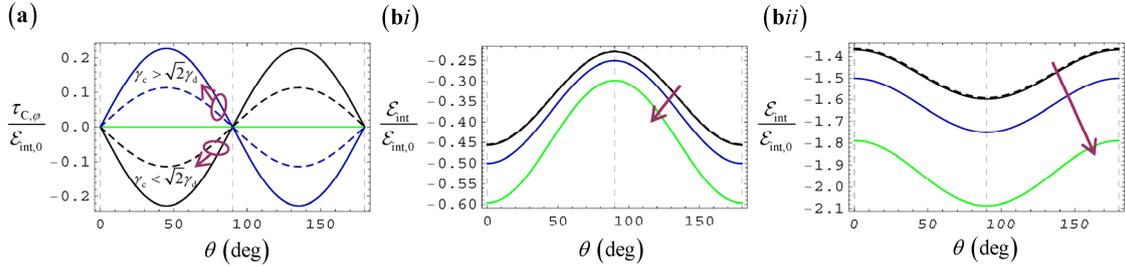

**Fig. 5 a)** Normalized Casimir torque acting on a Kramers two-level atom as a function of the orientation $\theta$ of the atom symmetry axis with respect to the *z*-direction. The atom transition frequency is $\omega_0 = 0.1\omega_{\text{sp}}$. Solid black line: $\gamma_c = 0$; Dashed black line: $\gamma_c = \gamma_d$; Green line: $\gamma_c = \sqrt{2}\gamma_d$; Dashed blue line: $\gamma_c = \sqrt{3}\gamma_d$; Solid blue line: $\gamma_c = 2\gamma_d$. The stable equilibrium orientation is $\theta = 0°, 180°$ for the black lines (unbroken rotational symmetry) and $\theta = 90°$ for the blue lines (spontaneously broken symmetry). **b*i*)** Normalized Casimir free energy as a function of the orientation of the atom symmetry axis for $\gamma_c = 0$ and $\omega_0 = 0.1\omega_{\text{sp}}$. Dashed black line: zero temperature limit; Black line: $\omega_T = \omega_0$; Blue line: $\omega_T = 2\omega_0$; Green line: $\omega_T = 3\omega_0$. The arrow indicates the direction of increasing temperature. **b*ii*)** Similar to b*i*) but for $\gamma_c = 2\gamma_d$.

## VI. Summary

The standard model of a two-level atom violates the time-reversal symmetry. To circumvent this deficiency, I developed an alternative model wherein the atom is characterized by two Kramers pairs of ground and excited states. In general, the two excited states can be coupled through dipolar-type transitions to any of the ground states. It was shown that in general it is impossible to eliminate crossed dipole transitions due to the spin-orbit interaction component of the atom Hamiltonian. It was demonstrated that the atom (parametric) electric polarizability for a definite ground-state is generically nonreciprocal.

The proposed model was applied to study Casimir-Polder forces when the Kramers two-level atom interacts with either a nonreciprocal or with a reciprocal environment. It was found that for a nonreciprocal environment the ground-state degeneracy is lifted by the interactions with the electromagnetic vacuum. In particular, the non-reciprocity of the substrate may induce atomic energy shifts that are reminiscent of the Zeeman effect. In contrast, for a reciprocal system, the Casimir-Polder physics is independent of the atomic ground state.

I investigated the stable equilibrium positions of a Kramers two-level system with symmetry of rotation about some axis when it is placed nearby a metal surface. It was shown that the Casimir torque acts to reorient the symmetry axis of the atom. Surprisingly, the stable equilibrium orientation does not always correspond to the alignment of the substrate and atom symmetry axes. When the chiral-type dipolar transitions dominate, the quantum fluctuations act to reorient the atom symmetry axis in such a way that it becomes parallel to the metallic surface. Thus, the quantum vacuum fluctuations lead to a stable ground state with a spontaneously broken rotational symmetry. The present theory highlights the richness of physics that can emerge due to the degeneracy of atomic states.

**Acknowledgements:** This work is supported in part by the IET under the A F Harvey Engineering Research Prize, by Fundação para a Ciência e a Tecnologia grant number PTDC/EEI-TEL/4543/2014 and by Instituto de Telecomunicações under project UID/EEA/50008/2019.

## Appendix A: Transformation of $\gamma_d$ and $\gamma_c$ under a change of basis

In this Appendix, I study how a change of basis affects the (vector) amplitudes of the transition dipole moment vector elements $\gamma_d$ and $\gamma_c$.

A change of basis is completely characterized by the coefficients $a_i, b_i$ ($i$=1,2) such that the new basis elements are

$$|e_1'\rangle = a_1|e_1\rangle + a_2|e_2\rangle, \qquad |g_1'\rangle = b_1|g_1\rangle + b_2|g_2\rangle, \tag{A1}$$

with $|a_1|^2 + |a_2|^2 = 1 = |b_1|^2 + |b_2|^2$. The remaining elements of the basis are obtained from the constraints $|e_2'\rangle = \mathcal{T}|e_1'\rangle$ and $|g_2'\rangle = \mathcal{T}|g_1'\rangle$ [Eq. (1)]:

$$|e_2'\rangle = -a_2^*|e_1\rangle + a_1^*|e_2\rangle, \qquad |g_2'\rangle = -b_2^*|g_1\rangle + b_1^*|g_2\rangle. \tag{A2}$$

In the new basis, $\gamma_{d'} \equiv \langle g_1'|\hat{\mathbf{p}}|e_1'\rangle$ and $\gamma_{c'} \equiv \langle g_1'|\hat{\mathbf{p}}|e_2'\rangle$ satisfy

$$\gamma_{d'} = a_1 b_1^* \gamma_d + a_2 b_2^* \gamma_d^* + a_2 b_1^* \gamma_c - a_1 b_2^* \gamma_c^*, \tag{A3a}$$

$$\gamma_{c'} = \left(-a_2 b_1 \gamma_d^* + a_1 b_2 \gamma_d + a_1 b_1 \gamma_c^* + a_2 b_2 \gamma_c\right)^*. \tag{A3b}$$

The relations $\gamma_d = \gamma_{g1,e1} = \gamma^*_{g2,e2}$ and $\gamma_c = \gamma_{g1,e2} = -\gamma^*_{g2,e1}$ [Eq. (3)] were used. Furthermore, a ground state of the form $|\psi_0\rangle = c_1|g_1\rangle + c_2|g_2\rangle$ is expressed in the new basis as $|\psi_0\rangle = c'_1|g'_1\rangle + c'_2|g'_2\rangle$ with $c'_i$ given by:

$$c'_1 = b^*_1 c_1 + b^*_2 c_2, \qquad c'_2 = -b_2 c_1 + b_1 c_2. \tag{A4}$$

It can be checked that the polarizability tensor (5) is invariant under a change of basis, i.e., under a transformation of the type $\gamma_d \to \gamma_{d'}$, $\gamma_c \to \gamma_{c'}$ and $c_i \to c'_i$, as it should be.

Suppose that some (primed) basis (with transition dipole moments $\gamma_{d'}, \gamma_{c'}$) is given. Is it possible to switch to another (unprimed basis) where $\gamma_c = 0$? Clearly, if this is possible Eq. (A3a) implies that $\gamma_{d'} = a_1 b^*_1 \gamma_d + a_2 b^*_2 \gamma^*_d$. Combining this equation with its complex conjugate one gets

$$\left(|a_1 b_1|^2 - |a_2 b_2|^2\right)\gamma_d = a^*_1 b_1 \gamma_{d'} - a_2 b^*_2 \gamma^*_{d'}. \tag{A5}$$

On the other hand, from Eq. (A3b) the relation $\gamma^*_{c'} = a_1 b_2 \gamma_d - a_2 b_1 \gamma^*_d$ must be also satisfied. The leading coefficient on the left-hand side of Eq. (A5) is necessarily non-zero when $\gamma_{d'}, \gamma^*_{d'}$ are linearly independent. In such a case, $\gamma_d$ is a linear combination of $\gamma_{d'}, \gamma^*_{d'}$, and the crossed transition element ($\gamma_c$) can be set identical to zero only if $\gamma_{c'}$ is a linear combination of $\gamma_{d'}, \gamma^*_{d'}$.

Furthermore, an identity analogous to Eq. (A3b) also holds true with the primed and unprimed symbols interchanged. Thus, when $\gamma_{d'}, \gamma^*_{d'}$ are linearly dependent it follows that $\gamma_c$ can be set equal to zero only if $\gamma_{d'}, \gamma_{c'}, \gamma^*_{c'}$ are linearly dependent.

In summary, in order that the crossed dipole moment ($\gamma_c$) can be set equal to zero with suitable basis change it is *necessary* that *i)* if $\gamma_{d'}, \gamma_{d'}^*$ are linearly independent $\gamma_{c'}$ must be a linear combination of $\gamma_{d'}, \gamma_{d'}^*$. *ii)* if $\gamma_{d'}, \gamma_{d'}^*$ are linearly dependent then $\gamma_{d'}, \gamma_{c'}, \gamma_{c'}^*$ must be linearly dependent.

## Appendix B: The atomic polarizability

In this Appendix, the free-space atomic polarizability is calculated using a linear response approximation. The Hamiltonian of the quantum system under the action of a classical electric field is:

$$\hat{H} = \hat{H}_{at} + \hat{H}_{int}, \tag{B1}$$

with $\hat{H}_{at} = \sum_i E_i |i\rangle\langle i|$ and $\hat{H}_{int} = -\hat{\mathbf{p}} \cdot \mathbf{E}$. It is assumed that the quantum system is formed by two pairs of degenerate energy states (Fig. 1a). When the stationary states have a vanishing dipole moment, the dipole moment operator $\hat{\mathbf{p}} = \hat{\mathbf{p}}^- + \hat{\mathbf{p}}^+$ is determined by Eq. (4).

The time-dynamics of $\hat{\mathbf{p}}^-$ is determined by the Heisenberg equation of motion $\frac{d\hat{\mathbf{p}}^-}{dt} = \frac{i}{\hbar}[\hat{H}, \hat{\mathbf{p}}^-]$, which is equivalent to

$$\frac{d}{dt}\hat{\mathbf{p}}^- = -i\omega_0 \hat{\mathbf{p}}^- - \frac{i}{\hbar}[\hat{\mathbf{p}}^+ \cdot \mathbf{E}, \hat{\mathbf{p}}^-], \tag{B2}$$

with $\omega_0 = (E_e - E_g)/\hbar$. It is assumed that the atomic state is of the form $|\psi_0\rangle = c_1|g_1\rangle + c_2|g_2\rangle$ with $|c_1|^2 + |c_2|^2 = 1$. The equation is linearized by replacing $[\hat{\mathbf{p}}^+ \cdot \mathbf{E}, \hat{\mathbf{p}}^-]$ by its expectation calculated at initial time:

$\left[ \hat{\mathbf{p}}^+ \cdot \mathbf{E}, \hat{\mathbf{p}}^- \right] \rightarrow \langle \psi_0 | \left[ \hat{\mathbf{p}}^+_{t=0} \cdot \mathbf{E}, \hat{\mathbf{p}}^-_{t=0} \right] | \psi_0 \rangle = -\langle \psi_0 | \hat{\mathbf{p}}^-_{t=0} \otimes \hat{\mathbf{p}}^+_{t=0} | \psi_0 \rangle \cdot \mathbf{E}$, with $\hat{\mathbf{p}}^{\pm}_{t=0}$ given by Eqs. (4b) and (4c). Denoting $\langle \hat{\mathbf{p}}^- \otimes \hat{\mathbf{p}}^+ \rangle_0 \equiv \langle \psi_0 | \hat{\mathbf{p}}^-_{t=0} \otimes \hat{\mathbf{p}}^+_{t=0} | \psi_0 \rangle$, the Heisenberg equation reduces to:

$$\frac{d}{dt}\hat{\mathbf{p}}^- = -i\omega_0 \hat{\mathbf{p}}^- + \frac{i}{\hbar} \langle \hat{\mathbf{p}}^- \otimes \hat{\mathbf{p}}^+ \rangle_0 \cdot \mathbf{E}, \tag{B3}$$

The equation can be explicitly solved for a time harmonic electric field ($\mathbf{E}(t) = \mathbf{E}_\omega e^{-i\omega t} + c.c.$):

$$\hat{\mathbf{p}}^- = \hat{\mathbf{p}}^-_{t=0} + \frac{1}{\hbar} \frac{\langle \hat{\mathbf{p}}^- \otimes \hat{\mathbf{p}}^+ \rangle_0}{\omega_0 - \omega} \cdot \mathbf{E}_\omega e^{-i\omega t} + \frac{1}{\hbar} \frac{\langle \hat{\mathbf{p}}^- \otimes \hat{\mathbf{p}}^+ \rangle_0}{\omega_0 + \omega^*} \cdot \mathbf{E}^*_\omega e^{+i\omega^* t}. \tag{B4}$$

Since $\hat{\mathbf{p}}^+ = (\hat{\mathbf{p}}^-)^\dagger$ and $\langle \hat{\mathbf{p}}_{t=0} \rangle = 0$ it can be easily verified that the expectation of the dipole moment can be written as $\langle \hat{\mathbf{p}} \rangle = \varepsilon_0 \boldsymbol{\alpha} \cdot \mathbf{E}_\omega e^{-i\omega t} + c.c.$, with the polarizability tensor $\boldsymbol{\alpha}$ given by:

$$\boldsymbol{\alpha}(\omega) = \frac{1}{\varepsilon_0 \hbar} \frac{\langle \hat{\mathbf{p}}^- \otimes \hat{\mathbf{p}}^+ \rangle_0}{\omega_0 - \omega} + \frac{1}{\varepsilon_0 \hbar} \frac{\langle \hat{\mathbf{p}}^- \otimes \hat{\mathbf{p}}^+ \rangle_0^*}{\omega_0 + \omega}. \tag{B5}$$

Using Eq. (4) one obtains:

$$\langle \hat{\mathbf{p}}^- \otimes \hat{\mathbf{p}}^+ \rangle_0 = \left[ \left( \boldsymbol{\gamma}_d \otimes \boldsymbol{\gamma}^*_d + \boldsymbol{\gamma}_c \otimes \boldsymbol{\gamma}^*_c \right) |c_1|^2 + \left( \boldsymbol{\gamma}^*_d \otimes \boldsymbol{\gamma}_d + \boldsymbol{\gamma}^*_c \otimes \boldsymbol{\gamma}_c \right) |c_2|^2 \right] \\ + \left[ \left( -\boldsymbol{\gamma}_d \otimes \boldsymbol{\gamma}_c + \boldsymbol{\gamma}_c \otimes \boldsymbol{\gamma}_d \right) c^*_1 c_2 + \left( \boldsymbol{\gamma}^*_d \otimes \boldsymbol{\gamma}^*_c - \boldsymbol{\gamma}^*_c \otimes \boldsymbol{\gamma}^*_d \right) c^*_2 c_1 \right]. \tag{B6}$$

From here, it is seen that $\boldsymbol{\alpha}(\omega)$ is given by

$$\boldsymbol{\alpha} = \frac{1}{\varepsilon_0 \hbar} \left\{ |c_1|^2 \left( \frac{\boldsymbol{\gamma}_d \otimes \boldsymbol{\gamma}_d^* + \boldsymbol{\gamma}_c \otimes \boldsymbol{\gamma}_c^*}{\omega_0 - \omega} + \frac{\boldsymbol{\gamma}_d^* \otimes \boldsymbol{\gamma}_d + \boldsymbol{\gamma}_c^* \otimes \boldsymbol{\gamma}_c}{\omega_0 + \omega} \right) \right.$$
$$\left. + |c_2|^2 \left( \frac{\boldsymbol{\gamma}_d^* \otimes \boldsymbol{\gamma}_d + \boldsymbol{\gamma}_c^* \otimes \boldsymbol{\gamma}_c}{\omega_0 - \omega} + \frac{\boldsymbol{\gamma}_d \otimes \boldsymbol{\gamma}_d^* + \boldsymbol{\gamma}_c \otimes \boldsymbol{\gamma}_c^*}{\omega_0 + \omega} \right) \right\} \quad \text{(B7)}$$
$$+ \frac{1}{\varepsilon_0 \hbar} i \tilde{\boldsymbol{\Omega}} \times \mathbf{1} \left( \frac{1}{\omega_0 - \omega} - \frac{1}{\omega_0 + \omega} \right)$$

with $\tilde{\boldsymbol{\Omega}} = -i\left( \boldsymbol{\gamma}_d \times \boldsymbol{\gamma}_c c_1^* c_2 - \boldsymbol{\gamma}_d^* \times \boldsymbol{\gamma}_c^* c_2^* c_1 \right)$. Writing the polarizability as a sum of symmetric and anti-symmetric tensors and taking into account that $|c_1|^2 + |c_2|^2 = 1$, one obtains Eq. (5) of the main text.

## Appendix C: The interaction tensor

The interaction tensor $\mathbf{C}^{\text{int}}$ introduced in Sect. IV can be expressed in terms of the scattering part of the system Green's function [25, 42]. The Green's function $\overline{\mathbf{G}}(\mathbf{r}, \mathbf{r}_0)$ (6×6 tensor) is defined here as the solution of [42, 51]

$$\hat{N} \cdot \overline{\mathbf{G}} = \omega \mathbf{M}(\mathbf{r}) \cdot \overline{\mathbf{G}} + i \mathbf{1}_{6 \times 6} \delta(\mathbf{r} - \mathbf{r}_0). \quad \text{(C1)}$$

with the differential operator $\hat{N}$ and the material matrix $\mathbf{M}$ given by

$$\hat{N} = \begin{pmatrix} 0 & i\nabla \times \mathbf{1}_{3\times 3} \\ -i\nabla \times \mathbf{1}_{3\times 3} & 0 \end{pmatrix}, \quad \text{and} \quad \mathbf{M} = \begin{pmatrix} \varepsilon_0 \overline{\varepsilon}(\mathbf{r}, \omega) \mathbf{1}_{3\times 3} & 0 \\ 0 & \mu_0 \mathbf{1}_{3\times 3} \end{pmatrix}. \quad \text{(C2)}$$

In the above, $\mathbf{1}_{3\times 3}$ is the identity matrix of dimension three and $\overline{\varepsilon}$ is the relative permittivity tensor. The total field radiated by a classical electric dipole can be expressed in terms of the Green function as $\mathbf{E} = -i\omega \mathbf{G}_{ee} \cdot \mathbf{p}$ with $\mathbf{G}_{ee}$ the 3×3 tensor obtained from the 3×3 upper-left block of $\overline{\mathbf{G}}$ and $\mathbf{p}$ the electric dipole moment.

Let $\mathbf{E}_{loc}(\mathbf{r}_0) \equiv \mathbf{E} - \mathbf{E}^{self}$ be the local field at the position of the dipole, i.e., the difference between the total electric field and the field radiated by the dipole alone in free-space ($\mathbf{E}^{self}$). The local field may be written as $\mathbf{E}_{loc} = -i\omega \mathbf{G}_{ee}^{scat}(\mathbf{r}_0, \mathbf{r}_0) \cdot \mathbf{p}$ where $\mathbf{G}_{ee}^{scat}$ is the scattering part of the Green function (total Green function with the self-field excluded). The interaction tensor $\mathbf{C}^{int}$ is by definition $\mathbf{C}_{int} = -i\omega\varepsilon_0 \mathbf{G}_{ee}^{scat}(\mathbf{r}_0, \mathbf{r}_0)$ so that the local field is related to the electric dipole moment as:

$$\mathbf{E}_{loc} = \mathbf{C}_{int} \cdot \frac{\mathbf{p}}{\varepsilon_0}. \tag{C3}$$

Supposing that the region above the material substrate in Fig. 1b is a vacuum, the interaction tensor is given by [42, 52]

$$\mathbf{C}_{int} = \frac{1}{(2\pi)^2} \iint dk_x dk_y \left[ \mathbf{1}_t + \frac{i}{\gamma_0} \hat{\mathbf{z}} \otimes \mathbf{k}_\parallel \right] \cdot \mathbf{R}(\omega, k_x, k_y) \cdot \left[ i\gamma_0 \mathbf{k}_\parallel \otimes \hat{\mathbf{z}} + \left(\frac{\omega^2}{c^2}\right) \mathbf{1}_t - \mathbf{k}_\parallel \otimes \mathbf{k}_\parallel \right] \frac{e^{-2\gamma_0 d}}{2\gamma_0}, \tag{C4}$$

where $\mathbf{1}_t = \hat{\mathbf{x}} \otimes \hat{\mathbf{x}} + \hat{\mathbf{y}} \otimes \hat{\mathbf{y}}$ is the transverse identity tensor, $\mathbf{k}_\parallel = k_x \hat{\mathbf{x}} + k_y \hat{\mathbf{y}}$, $\gamma_0 = -i\sqrt{(\omega/c)^2 - \mathbf{k}_\parallel \cdot \mathbf{k}_\parallel}$ and $d$ is the distance of the dipole with respect to the interface (plane $z=0$). Here, $\mathbf{R}(\omega, k_x, k_y)$ is a 2×2 matrix that links the tangential components of the reflected and incident fields, $\begin{pmatrix} E_x^{ref} \\ E_y^{ref} \end{pmatrix} = \mathbf{R}(\omega, k_x, k_y) \cdot \begin{pmatrix} E_x^{inc} \\ E_y^{inc} \end{pmatrix}$, for plane wave incidence on the substrate [52]. In the quasi-static limit, the term $\omega^2/c^2 \mathbf{1}_t$ in Eq. (C4) can be dropped, and within this approximation $C_{int,zz} = C_{int,xx} + C_{int,yy}$.

The matrix $\mathbf{R}$ can be written in terms of the matrix $\widetilde{\mathbf{R}} = \begin{pmatrix} R_{pp} & R_{ps} \\ R_{sp} & R_{ss} \end{pmatrix}$ that relates the amplitudes of incident and reflected $p$ and $s$ polarized waves. A straightforward analysis shows that:

$$\mathbf{R} = \frac{1}{k_\parallel} \begin{pmatrix} k_x \frac{k_{z0}}{k_0} & -k_y \\ k_y \frac{k_{z0}}{k_0} & k_x \end{pmatrix} \cdot \widetilde{\mathbf{R}} \cdot \frac{k_0}{k_\parallel k_{z0}} \begin{pmatrix} k_x & k_y \\ -k_y \frac{k_{z0}}{k_0} & k_x \frac{k_{z0}}{k_0} \end{pmatrix}, \quad (C5)$$

with $k_{z0} = i\gamma_0$ and $k_0 = \omega/c$. For reciprocal systems $\widetilde{\mathbf{R}}(\omega, k_x, k_y) = \left[\widetilde{\mathbf{R}}(\omega, -k_x, -k_y)\right]^T$ (the corresponding reciprocity relation for the matrix $\mathbf{R}$ is cumbersome, and hence is omitted here).

When the $p$ and $s$ polarizations are uncoupled (e.g., for any isotropic dielectric substrate), the matrix $\widetilde{\mathbf{R}}$ is diagonal and the $\mathbf{R}$ matrix is given by:

$$\mathbf{R} = \frac{1}{k_x^2 + k_y^2} \begin{pmatrix} R_{pp} k_x^2 + R_{ss} k_y^2 & (R_{pp} - R_{ss}) k_x k_y \\ (R_{pp} - R_{ss}) k_x k_y & R_{pp} k_y^2 + R_{ss} k_x^2 \end{pmatrix}. \quad (C6)$$

Here, $R_{pp}, R_{ss}$ are the standard (tangential electric field) reflection coefficients for $p$ and $s$ polarizations (see Ref. [52]). For example, for a perfect electric conductor $R_{pp} = R_{ss} = -1$ and $\mathbf{R} = -\mathbf{1}_t$.

## Appendix D: $\mathbf{C}_{int}$ for a metal half-space

Here, I obtain an explicit formula for $\mathbf{C}_{int}$ for a metal half-space using a quasi-static approximation.

For lossless systems, the Green function can be expanded in terms of the normal modes ($\mathbf{f}_{n\mathbf{k}} = \begin{pmatrix} \mathbf{E}_{n\mathbf{k}} & \mathbf{H}_{n\mathbf{k}} \end{pmatrix}^T$) as [51, 53, 54]:

$$-i\omega\overline{\mathbf{G}} = \sum_{\omega_{n\mathbf{k}}>0} \frac{\omega_{n\mathbf{k}}}{2} \left( \frac{1}{\omega_{n\mathbf{k}} - \omega} \mathbf{f}_{n\mathbf{k}}(\mathbf{r}) \otimes \mathbf{f}^*_{n\mathbf{k}}(\mathbf{r}') + \frac{1}{\omega_{n\mathbf{k}} + \omega} \mathbf{f}^*_{n\mathbf{k}}(\mathbf{r}) \otimes \mathbf{f}_{n\mathbf{k}}(\mathbf{r}') \right) - \mathbf{M}_{\infty}^{-1} \delta(\mathbf{r} - \mathbf{r}'),$$

(D1)

where $\mathbf{M}_{\infty} = \mathbf{M}(\omega = \infty)$ and $\omega_{n\mathbf{k}}$ are the resonant frequencies. The modes are normalized as:

$$\frac{1}{2} \int d^3\mathbf{r}\, \mathbf{f}^*_{n\mathbf{k}} \cdot \frac{\partial}{\partial \omega}[\omega \mathbf{M}]_{\omega=\omega_{n\mathbf{k}}} \cdot \mathbf{f}_{n\mathbf{k}} = 1.$$

(D2)

As discussed in Appendix C, the electric field radiated by an electric dipole with dipole moment $\mathbf{p}$ is given by $\mathbf{E} = \overline{\mathcal{G}} \cdot \mathbf{p}$ with $\overline{\mathcal{G}} = -i\omega\overline{\mathbf{G}}_{ee}$. Evidently, the tensor $\overline{\mathcal{G}}$ has a decomposition analogous to $\overline{\mathbf{G}}$:

$$\overline{\mathcal{G}} = \overline{\mathcal{G}}^+ + \overline{\mathcal{G}}^- - \mathbf{1}_{3\times 3} \frac{1}{\varepsilon_0 \varepsilon_\infty} \delta(\mathbf{r} - \mathbf{r}').$$

(D3a)

$$\overline{\mathcal{G}}^+ = \sum_{\omega_{n\mathbf{k}}>0} \frac{\omega_{n\mathbf{k}}}{2} \frac{1}{\omega_{n\mathbf{k}} - \omega} \mathbf{E}_{n\mathbf{k}}(\mathbf{r}) \otimes \mathbf{E}^*_{n\mathbf{k}}(\mathbf{r}'), \qquad \overline{\mathcal{G}}^- = \sum_{\omega_{n\mathbf{k}}>0} \frac{\omega_{n\mathbf{k}}}{2} \frac{1}{\omega_{n\mathbf{k}} + \omega} \mathbf{E}^*_{n\mathbf{k}}(\mathbf{r}) \otimes \mathbf{E}_{n\mathbf{k}}(\mathbf{r}').$$

(D3b)

When the emitter is placed in the vicinity of the metal half-space the retardation effects due to the finite speed of light are negligible. In these conditions, it is helpful to use a quasi-static approximation such that the complex field amplitudes satisfy $\mathbf{E} \approx -\nabla\phi$ and $\mathbf{H} \approx 0$. In the quasi-static limit, the modes are surface plasmon polaritons described by $\mathbf{E}_{n\mathbf{k}} = -\nabla\phi_{\mathbf{k}}$ and $\phi_{\mathbf{k}} = A_{\mathbf{k}} e^{i\mathbf{k}_\parallel \cdot \mathbf{r}} e^{-k_\parallel |z|}$ [42, 53]. Assuming that the metal permittivity is $\varepsilon(\omega) = 1 - 2\omega_{sp}^2 / \omega^2$, the resonance frequencies are $\omega_{\mathbf{k}} = \omega_{sp}$ [53]. Thus, the quasi-static approximation yields,

$$\overline{\mathcal{G}}^+ = \frac{\omega_{sp}}{2} \frac{1}{\omega_{sp} - \omega} \sum_{\omega_{n\mathbf{k}} > 0} \mathbf{E}_{n\mathbf{k}}(\mathbf{r}) \otimes \mathbf{E}_{n\mathbf{k}}^*(\mathbf{r}'), \tag{D4}$$

From Eq. (D2), the normalization coefficient is found to be $A_{\mathbf{k}} = \sqrt{\dfrac{1}{A_s \varepsilon_0 2 k_\parallel}}$, where $A_s$ is the area of the metallic surface [53]. Letting $A_s \to \infty$ it is found that:

$$\begin{aligned}\overline{\mathcal{G}}^+ &= \frac{\omega_{sp}}{2} \frac{1}{\omega_{sp} - \omega} \left[ \frac{1}{A_s} \sum_{\mathbf{k}_\parallel} \frac{1}{\varepsilon_0 2 k_\parallel} \vec{\nabla} e^{i\mathbf{k}_\parallel \cdot (\mathbf{r} - \mathbf{r}')} e^{-k_\parallel (|z| + |z'|)} \overleftarrow{\nabla}' \right] \\ &= \frac{\omega_{sp}}{2} \frac{1}{\omega_{sp} - \omega} \vec{\nabla} \left[ \frac{1}{(2\pi)^2} \iint d^2 \mathbf{k}_\parallel \frac{1}{\varepsilon_0 2 k_\parallel} e^{i\mathbf{k}_\parallel \cdot (\mathbf{r} - \mathbf{r}')} e^{-k_\parallel (|z| + |z'|)} \right] \overleftarrow{\nabla}'\end{aligned} \tag{D5}$$

This gives the closed-form result:

$$\overline{\mathcal{G}}^+ = \frac{\omega_{sp}}{2} \frac{1}{\omega_{sp} - \omega} \vec{\nabla} \left[ \frac{1}{4\pi\varepsilon_0} \frac{1}{\sqrt{(x-x')^2 + (y-y')^2 + (|z| + |z'|)^2}} \right] \overleftarrow{\nabla}'. \tag{D6}$$

The term $\overline{\mathcal{G}}^-(\omega)$ is given by a similar formula with $-\omega$ in the place of $\omega$. In the quasi-static approximation, the scattering part of the Green function can be identified with $\overline{\mathcal{G}}^+ + \overline{\mathcal{G}}^-$ and satisfies:

$$\overline{\mathcal{G}}^+ + \overline{\mathcal{G}}^- = \frac{\varepsilon(\omega) - 1}{\varepsilon(\omega) + 1} \vec{\nabla} \left[ \frac{1}{4\pi\varepsilon_0} \frac{1}{\sqrt{(x-x')^2 + (y-y')^2 + (|z| + |z'|)^2}} \right] \overleftarrow{\nabla}'. \tag{D7}$$

Note that $\dfrac{\varepsilon(\omega) - 1}{\varepsilon(\omega) + 1} = \dfrac{\omega_{sp}^2}{\omega_{sp}^2 - \omega^2}$ for a lossless Drude model. Even though the previous analysis ignored the effect of loss, the final result [Eq. (D7)] can be readily extended to lossy systems simply by using the lossy permittivity function ($\varepsilon(\omega) = 1 - 2\omega_{sp}^2 / \omega(\omega + i\Gamma_m)$) in the formula. In the zero frequency limit $\frac{\varepsilon(\omega) - 1}{\varepsilon(\omega) + 1} \to 1$ and $\overline{\mathcal{G}}^+ + \overline{\mathcal{G}}^-$ gives precisely the field back-scattered by a PEC surface, i.e., the field created by an image dipole. Using $\mathbf{C}_{int} = -i\omega\varepsilon_0 \mathbf{G}_{ee}^{scat}(\mathbf{r}_0, \mathbf{r}_0)$ with $-i\omega \mathbf{G}_{ee}^{scat}(\mathbf{r}_0, \mathbf{r}_0) = \overline{\mathcal{G}}^+ + \overline{\mathcal{G}}^-$ one obtains the result of the main text [Eq. (14)].

## Appendix E: Effective Ground-state Hamiltonian

Here, I derive an effective ground-state Hamiltonian for the Kramers two-level atom using lowest order perturbation theory. When the substrate is reciprocal the effective Hamiltonian is always represented by a scalar (real-valued number).

The Hamiltonian of the system can be written as $\hat{H} = \hat{H}_0 + \hat{H}_{int}$ where $\hat{H}_0 = \hat{H}_{at} + \hat{H}_{EM}$ is the Hamiltonian for the non-interacting (free) atom and field, and $\hat{H}_{int} = -\hat{\mathbf{p}} \cdot \hat{\mathbf{E}}$ is the interaction term. The Hamiltonian can be represented by a matrix in a basis formed by the free-ground states $G = \{|g_1 0\rangle, |g_2 0\rangle\}$ ($|0\rangle$ represents the quantum vacuum) and by the free-field "excited" states, $E = \{|g_i 1_j\rangle, |e_i 1_j\rangle, ..., |g_i F\rangle, |e_i F\rangle\}$ where $F$ is a generic field state with one or more light quanta. In this basis, the Hamiltonian is represented by:

$$\hat{H} \to \begin{bmatrix} H_{GG} & H_{GE} \\ H_{EG} & H_{EE} \end{bmatrix}. \tag{E1}$$

Here, $H_{GG} = \left(\hat{H}_0 + \hat{H}_{int}\right)\big|_{GG}$ is a 2×2 matrix with elements $\langle g_m 0 | \hat{H}_0 + \hat{H}_{int} | g_n 0 \rangle$ (m, n=1,2), $H_{GE} = \hat{H}_{int}\big|_{GE}$ is a 2×∞ matrix, and so on. The energy eigenstates are the nontrivial solutions of:

$$\begin{bmatrix} H_{GG} - E\mathbf{1}_{2\times 2} & H_{GE} \\ H_{EG} & H_{EE} - E\mathbf{1} \end{bmatrix} \cdot \begin{pmatrix} \mathbf{c}_G \\ \mathbf{c}_E \end{pmatrix} = 0. \tag{E2}$$

For simplicity, the energy of the free-ground states is taken identical to zero ($\hat{H}_0\big|_{GG} = 0$). Then, since $\langle g_m 0 | \hat{H}_{int} | g_n 0 \rangle = 0$, it follows that $H_{GG} = \mathbf{0}_{2\times 2}$. Writing the free-excited states coefficients ($\mathbf{c}_E$) in terms of the free-ground states coefficients ($\mathbf{c}_G$), the secular equation reduces to

$$\left[ -E\mathbf{1}_{2\times 2} - H_{GE} \cdot (H_{EE} - E\mathbf{1})^{-1} \cdot H_{EG} \right] \cdot \mathbf{c}_G = 0. \tag{E3}$$

Thus, the ground-state of the interacting system is characterized by the effective Hamiltonian $H_{ef} = -H_{GE} \cdot (H_{EE} - E\mathbf{1})^{-1} \cdot H_{EG}$. Since $H_{GE} = \hat{H}_{int}\big|_{GE}$ and $H_{EG} = \hat{H}_{int}\big|_{EG}$ are proportional to the interaction strength, to lowest order perturbation theory it is possible to replace $H_{EE} - E\mathbf{1}$ by $\hat{H}_0\big|_{EE}$ (note that the ground state energy of the interacting system must be of the same order as that of the free fields, i.e., $E \approx 0$). Hence, the effective Hamiltonian is:

$$H_{ef} \approx -\hat{H}_{int}\big|_{GE} \cdot \left(\hat{H}_0\big|_{EE}\right)^{-1} \cdot \hat{H}_{int}\big|_{EG}. \tag{E4}$$

Let $h_{m,n} = \langle g_m 0 | \hat{H}_{ef} | g_n 0 \rangle$ (m, n=1,2) be a generic matrix element of $H_{ef}$. The electric field operator can be written as [51, 53, 55]

$$\hat{\mathbf{E}}(\mathbf{r}) = \sum_{\omega_i > 0} \sqrt{\frac{\hbar \omega_i}{2}} \left( \mathbf{E}_i(\mathbf{r}) \hat{a}_i + \mathbf{E}_i^*(\mathbf{r}) \hat{a}_i^\dagger \right). \tag{E5}$$

with $\mathbf{E}_i$ a generic field mode (normalized as in Eq. (D2) [51, 53]) associated with the angular frequency $\omega_i$, and $\hat{a}_i, \hat{a}_i^\dagger$ are the corresponding annihilation and creation operators satisfying standard commutation relations ($[\hat{a}_i, \hat{a}_i^\dagger] = 1$, etc). Then, a straightforward analysis shows that:

$$\begin{aligned} h_{m,n} &= -\sum_{i,\tilde{i}} \langle g_m 0 | \hat{H}_{\text{int}} | e_{\tilde{i}} 1_i \rangle \frac{1}{\hbar(\omega_0 + \omega_i)} \langle e_{\tilde{i}} 1_i | \hat{H}_{\text{int}} | g_n 0 \rangle \\ &= -\sum_{i,\tilde{i}} \frac{\hbar \omega_i}{2} \langle g_m | \hat{\mathbf{p}} \cdot \mathbf{E}_i(\mathbf{r}_0) | e_{\tilde{i}} \rangle \frac{1}{\hbar(\omega_0 + \omega_i)} \langle e_{\tilde{i}} | \hat{\mathbf{p}} \cdot \mathbf{E}_i^*(\mathbf{r}_0) | g_n \rangle \end{aligned} \tag{E6}$$

In the above, $\omega_0$ and $\mathbf{r}_0$ are the transition frequency and coordinates, respectively, of the two-level system. Taking into account that $\langle g_\alpha | \hat{\mathbf{p}} | g_\beta \rangle = 0$ and $\mathbf{1} = |e_1\rangle\langle e_1| + |e_2\rangle\langle e_2| + |g_1\rangle\langle g_1| + |g_2\rangle\langle g_2|$ one finds that:

$$h_{m,n} = -\varepsilon_0 \hbar \sum_i \frac{\omega_i}{2} \frac{1}{\omega_0 + \omega_i} \text{tr}\left\{ \mathbf{E}_i^* \mathbf{E}_i \cdot \mathbf{R}^{(m,n)} \right\}. \tag{E7}$$

where $\mathbf{R}^{(m,n)} = \frac{1}{\varepsilon_0 \hbar} \langle g_m | \hat{\mathbf{p}} \otimes \hat{\mathbf{p}} | g_n \rangle$.

The matrix element $h_{m,n}$ can be written in terms of the system Green's function defined in Appendix D. Using the modal expansion (D3) one readily sees that $h_{m,n} = -\varepsilon_0 \hbar \, \text{tr}\left\{ \overline{\mathcal{G}^-}(\mathbf{r}_0, \mathbf{r}_0; \omega_0) \cdot \mathbf{R}^{(m,n)} \right\}$. From (D3) it can be checked that for $\text{Re}\{\omega_0\} > 0$:

$$\begin{aligned} \mathcal{G}^-(\mathbf{r}_0, \mathbf{r}_0; \omega_0) &= \frac{1}{2\pi} \int_{-\infty}^{\infty} d\xi \frac{1}{\omega_0 - i\xi} (\mathcal{G}^- + \mathcal{G}^+)(\mathbf{r}_0, \mathbf{r}_0; i\xi) \\ &= \frac{1}{2\pi} \int_0^\infty d\xi \frac{1}{\omega_0 - i\xi} (\mathcal{G}^- + \mathcal{G}^+)(\mathbf{r}_0, \mathbf{r}_0; i\xi) + \frac{1}{\omega_0 + i\xi} \left[ (\mathcal{G}^- + \mathcal{G}^+)(\mathbf{r}_0, \mathbf{r}_0; i\xi) \right]^T \end{aligned}, \tag{E8}$$

where it was taken into account that $\left[\left(\mathcal{G}^- + \mathcal{G}^+\right)(\mathbf{r},\mathbf{r}_0;-i\xi)\right] = \left[\left(\mathcal{G}^- + \mathcal{G}^+\right)(\mathbf{r}_0,\mathbf{r};i\xi)\right]^T$.

Combining the previous equation with $h_{m,n} = -\varepsilon_0 \hbar \, \mathrm{tr}\left\{\overline{\mathcal{G}}^-(\omega_0) \cdot \mathbf{R}^{(m,n)}\right\}$ one finds after some manipulations that:

$$h_{m,n} = -\frac{\hbar}{2\pi}\int_0^\infty d\xi \, \mathrm{tr}\left\{\mathbf{C}_{\mathrm{int}}(i\xi) \cdot \left(\frac{\mathbf{R}^{(m,n)}}{\omega_0 - i\xi} + \frac{\mathbf{R}^{(m,n),T}}{\omega_0 + i\xi}\right)\right\}, \quad m,n = 1,2, \tag{E9}$$

with $\mathbf{C}_{\mathrm{int}}(i\xi) = \varepsilon_0 \left(\mathcal{G}^- + \mathcal{G}^+\right)(\mathbf{r}_0,\mathbf{r}_0;i\xi)$ the interaction tensor of Appendix C. Note that the above formula holds true even when the material substrate is dissipative, as a lossy system can be regarded as the limit of a sequence of lossless systems [26].

The energy expectation for a generic ground state $|\psi_0\rangle = c_1|g_1 0\rangle + c_2|g_2 0\rangle$ is $\langle \psi_0 | H_{\mathrm{ef}} | \psi_0 \rangle = \sum_{m,n} c_m^* c_n h_{m,n}$. From Eq. (B5) the atomic polarizability can be expressed as

$$\boldsymbol{\alpha}(\omega;|\psi_0\rangle) = \sum_{m,n} c_m^* c_n \frac{\mathbf{R}^{(m,n)}}{\omega_0 - \omega} + \frac{1}{\varepsilon_0 \hbar}\frac{\mathbf{R}^{(m,n),T}}{\omega_0 + \omega}$$ (note that $\langle \hat{\mathbf{p}}^- \otimes \hat{\mathbf{p}}^- \rangle_0 = \langle \hat{\mathbf{p}}^- \otimes \hat{\mathbf{p}}^+ \rangle_0$ and $\langle \hat{\mathbf{p}}^- \otimes \hat{\mathbf{p}}^+ \rangle_0^* = \langle \hat{\mathbf{p}}^- \otimes \hat{\mathbf{p}}^+ \rangle_0^T$; the latter result is a consequence of Eq. (B6)). Using this formula for the polarizability, it is simple to verify that the energy expectation obtained from the effective Hamiltonian is exactly coincident with the Casimir interaction energy determined by Eq. (11): $\langle \psi_0 | H_{\mathrm{ef}} | \psi_0 \rangle = \mathcal{E}_{\mathrm{int}}(|\psi_0\rangle)$.

The matrices $\mathbf{R}^{(m,n)}$ can be explicitly evaluated as:

$$\mathbf{R}^{(1,1)} = \frac{1}{\hbar\varepsilon_0}\left(\boldsymbol{\gamma}_\mathrm{d} \otimes \boldsymbol{\gamma}_\mathrm{d}^* + \boldsymbol{\gamma}_\mathrm{c} \otimes \boldsymbol{\gamma}_\mathrm{c}^*\right), \qquad \mathbf{R}^{(2,2)} = \frac{1}{\hbar\varepsilon_0}\left(\boldsymbol{\gamma}_\mathrm{d}^* \otimes \boldsymbol{\gamma}_\mathrm{d} + \boldsymbol{\gamma}_\mathrm{c}^* \otimes \boldsymbol{\gamma}_\mathrm{c}\right), \tag{E10a}$$

$$\mathbf{R}^{(1,2)} = \frac{1}{\hbar\varepsilon_0}\left(-\boldsymbol{\gamma}_\mathrm{d} \otimes \boldsymbol{\gamma}_\mathrm{c} + \boldsymbol{\gamma}_\mathrm{c} \otimes \boldsymbol{\gamma}_\mathrm{d}\right), \qquad \mathbf{R}^{(2,1)} = \frac{1}{\hbar\varepsilon_0}\left(\boldsymbol{\gamma}_\mathrm{d}^* \otimes \boldsymbol{\gamma}_\mathrm{c}^* - \boldsymbol{\gamma}_\mathrm{c}^* \otimes \boldsymbol{\gamma}_\mathrm{d}^*\right). \tag{E10b}$$

The matrices $\mathbf{R}^{(1,2)}$ and $\mathbf{R}^{(2,1)}$ are anti-symmetric. Thereby, when the substrate is reciprocal ($\mathbf{C}_{int}$ is symmetric) it is necessary that $\text{tr}\left\{\mathbf{C}_{int}(i\xi)\cdot\mathbf{R}^{(m,n)}\right\} = 0 = \text{tr}\left\{\mathbf{C}_{int}(i\xi)\cdot\mathbf{R}^{(m,n),T}\right\}$ for $(m,n) = (1,2)$ and $(n,m) = (2,1)$ (as previously mentioned, the trace of the product of symmetric and anti-symmetric tensors vanishes). This implies that $h_{12} = 0 = h_{21}$. Furthermore, since the symmetric parts of $\mathbf{R}^{(1,1)}$ and $\mathbf{R}^{(2,2)}$ are identical, one has $h_{11} = h_{22}$. In conclusion, when the substrate is reciprocal the effective Hamiltonian is a scalar $H_{ef} = \mathcal{E}_{int}\mathbf{1}_{2\times 2}$.

In contrast, for a nonreciprocal substrate $H_{ef}$ has a nontrivial structure and in general a free-ground state of the form $|\psi_0\rangle = c_1|g_1 0\rangle + c_2|g_2 0\rangle$ is not an eigenstate of $H_{ef}$. When the Kramers pairs are uncoupled ($\gamma_c = 0$) the matrices $\mathbf{R}^{(1,2)}$ and $\mathbf{R}^{(2,1)}$ vanish and therefore $H_{ef}$ is a diagonal matrix. In this case, the ground state of the interacting system is in general non-degenerate and is either $|g_1 0\rangle$ or $|g_2 0\rangle$.

# References

[1] R. Feynman, R. Leighton, and M. Sands, The Feynman lectures on physics (California Institute of Technology, 1963).
[2] M. G. Silveirinha, "Time-reversal symmetry in Antenna Theory", *Symmetry*, **11**, 486, (2019).
[3] B. A. Bernervig, T. Hughes, *Topological Insulators and Topological Superconductors*, Princeton University Press, 2013.
[4] Shun-Qing Shen, *Topological Insulators,* (Series in Solid State Sciences,174), Springer, Berlin, 2012 (p. 98).
[5] M. G. Silveirinha, "PTD Symmetry Protected Scattering Anomaly in Optics", *Phys. Rev. B*, **95**, 035153, (2017).
[6] H. G. B. Casimir and D. Polder, "The Influence of Retardation on the London-van der Waals Forces", *Phys. Rev.* **73**, 360 (1948).

[7] H. B. G. Casimir, "On the attraction between two perfectly conducting plates", *Proc. K. Ned. Akad. Wet.* **51**, 791 (1948).

[8] E. M. Lifshitz, "The Theory of Molecular Attractive Forces between Solids", *Sov. Phys. JETP* **2**, 73 (1956).

[9] I. E. Dzyaloshinski, E. M. Lifshitz, and L. P. Pitaevski, "The general theory of van der Waals forces", *Adv. Phys.* **10**, 165 (1965).

[10] S. A. H. Gangaraj, G. W. Hanson, M. Antezza, and M. G. Silveirinha, "Spontaneous lateral atomic recoil force close to a photonic topological material", *Phys. Rev. B*, **97**, 201108(R), (2018).

[11] S. A. H. Gangaraj, M. G. Silveirinha, G. W. Hanson, M. Antezza, and F. Monticone, "Optical torque on a two-level system near a strongly nonreciprocal medium", *Phys. Rev. B*, **98**, 125146, (2018).

[12] Q.-D. Jiang, F. Wilczek, "Quantum Atmospherics for Materials Diagnosis", *Phys. Rev. B*, **99**, 201104(R), (2019).

[13] S. Fuchs, F. Lindel, R. V. Krems, G. W. Hanson, M. Antezza, and S. Y. Buhmann, "Casimir-Lifshitz force for nonreciprocal media and applications to photonic topological insulators", *Phys. Rev. A* **96**, 062505, (2017).

[14] S. Fuchs, J. A. Crosse, and S. Y. Buhmann, "Casimir-Polder shift and decay rate in the presence of nonreciprocal media", *Phys. Rev. A*, **95**, 023805, (2017).

[15] F. Lindel, G. W. Hanson, M. Antezza, and S. Y. Buhmann, "Inducing and controlling rotation on small objects using photonic topological materials", *Phys. Rev. B*, **98**, 144101, (2018).

[16] S. Y. Buhmann, V. N. Marachevsky, and S. Scheel, "Charge-parity-violating effects in Casimir-Polder potentials", *Phys. Rev. A*, **98**, 022510, (2018).

[17] P. W. Anderson, "More Is Different", *Science*, **177**, 393, (1972).

[18] J. van Wezel and J. van den Brink, "Spontaneous symmetry breaking in quantum mechanics", *Am. J. Phys.* **75** 635, (2007).

[19] S. Weinberg, "From BCS to the LHC", *Int.J.Mod.Phys.* **A23**, 1627, (2008).

[20] M. G. Silveirinha, "Spontaneous Parity-Time Symmetry Breaking in Moving Media", *Phys. Rev. A*, **90**, 013842, (2014).

[21] M. G. Silveirinha, "Theory of Quantum Friction", *New J. Phys.*, **16**, 063011 (1-29), (2014).

[22] C. L. Kane, E. J. Mele, "Quantum Spin Hall Effect in Graphene", *Phys. Rev. Lett.*, **95**, 226801, (2005).

[23] C. L. Kane, E. J. Mele, "$Z_2$ topological order and the quantum spin Hall effect", *Phys. Rev. Lett.*, **95**, 146802, (2005).

[24] M. Z. Hasan, C. L. Kane, "Colloquium: Topological insulators", *Rev. Mod. Phys.*, **82**, 3045, (2010).

[25] S. Y. Buhmann, L. Knoll, D.-G. Welsch, and H. T. Dung, "Casimir-Polder forces: A nonperturbative approach", *Phys. Rev. A*, **70**, 052117 (2004).

[26] M. G. Silveirinha, "Hidden Time-Reversal Symmetry in Dissipative Reciprocal Systems", *Opt. Express*, **27**, 14328, (2019).

[27] R. Loudon and S. M. Barnett, "Theory of the Linear Polarizability of a Two-Level Atom", *J. Phys. B* **39**, S555 (2006).

[28] C. Caloz, A. Alù, S. Tretyakov, D. Sounas, K. Achouri, and Z.-L. Deck-Léger, "Electromagnetic nonreciprocity," *Phys. Rev. Appl.*, **10**, 047001 (2018).

[29] P. Lodahl, S. Mahmoodian, S. Stobbe, A. Rauschenbeutel, P. Schneeweiss, J. Volz, H. Pichler, P. Zoller, "Chiral quantum optics", *Nature*, **541**, 473, (2017).

[30] C. Sayrin, C. Junge, R. Mitsch, B. Albrecht, D. O'Shea, P. Schneeweiss, J. Volz, and A. Rauschenbeutel, "Nanophotonic Optical Isolator Controlled by the Internal State of Cold Atoms", *Phys. Rev. X* **5**, 041036 (2015).

[31] I. Shomroni, S. Rosenblum, Y. Lovsky, O. Bechler, G. Guendelman, and B. Dayan, "All–optical routing of single photons by a one-atom switch controlled by a single photon", *Science*, **345**, 903 (2014).

[32] M. Scheucher, A. Hilico, E. Will, J. Volz, A. Rauschenbeutel, "Quantum optical circulator controlled by a single chirally coupled atom", *Science*, **354**, 1577–1580 (2016).

[33] P. Doyeux, S. A. H. Gangaraj, G. W Hanson, M. Antezza, "Giant interatomic energy-transport amplification with nonreciprocal photonic topological insulators", *Phys. Rev. Lett.*, **119**, 173901, (2017).

[34] C. A. Downing, J. C. López Carreño, F. P. Laussy, E. del Valle, and A. I. Fernández-Domínguez, "Quasichiral Interactions between Quantum Emitters at the Nanoscale", *Phys. Rev. Lett.*, **122**, 057401, (2019).

[35] C. Junge, D. O'Shea, J. Volz, A. Rauschenbeutel, "Strong coupling between single atoms and nontransversal photons" *Phys. Rev. Lett.*, **110**, 213604, (2013).

[36] R. Mitsch, C. Sayrin, B. Albrecht, P. Schneeweiss, and A. Rauschenbeutel, "Quantum state-controlled directional spontaneous emission of photons into a nanophotonic waveguide", *Nat. Commun.* **5**, 5713 (2014).

[37] I. Söllner, S. Mahmoodian, S. L. Hansen, L. Midolo, A. Javadi, G. Kiršanskė, T. Pregnolato, H. El-Ella, E. Hye Lee, J. D. Song, S. Stobbe, P. Lodahl, "Deterministic photon–emitter coupling in chiral photonic circuits", *Nat. Nanotechnol.* **10**, 775, (2015).

[38] S. Mittal, E. A. Goldschmidt, M. Hafezi, "A topological source of quantum light", *Nature*, **561**, 502, (2018).

[39] S. Barik, A. Karasahin, S. Mittal, E. Waks, M. Hafezi, "Chiral quantum optics using a topological resonator", arXiv:1906.11263.

[40] L. Onsager, "Reciprocal relations in irreversible processes I," *Phys. Rev.* **37**, 405 (1931).

[41] H. B. G. Casimir, "Reciprocity theorems and irreversible processes", *Proc. IEEE*, **51**, 1570, (1963).

[42] M. G. Silveirinha, S. A. H. Gangaraj, G. W. Hanson, M. Antezza, "Fluctuation-induced forces on an atom near a photonic topological material", *Phys. Rev. A*, **97**, 022509, (2018).

[43] L. D. Landau, E. M. Lifshitz, and L. P. Pitaevskii, *Electrodynamics of continuous media*, vol. 8 of Course on Theoretical Physics, (Butterworth-Heinemann, 2004).

[44] V. A. Parsegian, and G. H. Weiss, "On van der Waals interactions between macroscopic bodies having inhomogeneous dielectric susceptibilities", *Adv. Colloid Interface Sci.* **40**, 35, (1972).

[45] T. A. Morgado, M. G. Silveirinha, "Single-Interface Casimir Torque", *New J. Phys.*, **18**, 103030, (2016).

[46] D. A. T. Somers, J. L. Garrett, K. J. Palm, J. N. Munday, "Measurement of the Casimir torque", *Nature*, **564**, 386, (2018).

[47] S. Sanders, W. J. M. Kort-Kamp, D. A. R. Dalvit, A. Manjavacas, "Nanoscale transfer of angular momentum mediated by the Casimir torque", *Comm. Phys.*, **2**, 71 (2019).

[48] J. F. Babb, G. L. Klimchitskaya, and V. M. Mostepanenko, "Casimir-Polder interaction between an atom and a cavity wall under the influence of real conditions", *Phys. Rev. A*, **70**, 042901, (2004).

[49] G. L. Klimchitskaya, U. Mohideen, V. M. Mostepanenko, "The Casimir force between real materials: Experiment and theory", *Rev. Mod. Phys.* **81**, 1827, (2009).

[50] S. I. Maslovski, M. G. Silveirinha, "Casimir forces at the threshold of the Cherenkov effect", *Phys. Rev. A*, **84**, 062509 (2011).

[51] M. G. Silveirinha, "Modal expansions in dispersive material systems with application to quantum optics and topological photonics", chapter to appear in "Advances in Mathematical Methods for Electromagnetics", (edited by Paul Smith, Kazuya Kobayashi) IET, (available in arXiv:1712.04272).

[52] M. G. Silveirinha, "Optical instabilities and spontaneous light emission by polarizable moving matter", *Phys. Rev. X*, **4**, 031013, (2014).

[53] S. Lannebère, M. G. Silveirinha, "Negative Spontaneous Emission by a Moving Two-Level Atom", *J. Opt.*, **19**, 014004, (2017).

[54] M. G. Silveirinha, "Topological classification of Chern-type insulators by means of the photonic Green function", *Phys. Rev. B*, **97**, 115146, (2018).

[55] M. G. Silveirinha, "Topological Angular Momentum and Radiative Heat Transport in Closed Orbits", *Phys. Rev. B*, **95**, 115103, (2017).